\documentclass{elsart}
\usepackage{graphics}
\usepackage{color}
\usepackage{amssymb}
\usepackage{amsmath}
\usepackage{lscape}

\begin{document}
\begin{frontmatter}
\title{Relativistic formulation of Glauber theory for $A(e,e'p)$
reactions \thanksref{FWO}}
\author[Gent]{J. Ryckebusch},
%\email[]{jan.ryckebusch@rug.ac.be}
\author[Gent]{D. Debruyne},
\author[Gent]{P. Lava},
\author[Gent]{S. Janssen},
\author[Gent]{B. Van Overmeire} and
\author[Gent]{T. Van Cauteren}
\address[Gent]{Department of Subatomic and Radiation Physics, \\
Ghent University, Proeftuinstraat 86, B-9000 Gent, Belgium}
\thanks[FWO]{This work was supported by the Fund for Scientific Research - Flanders
under contract number G.0020.03 and the Research Council of Ghent
University.} 

\date{\today}

\begin{abstract}
At sufficiently large proton energies, Glauber multiple-scattering
theory offers good opportunities for describing the final state
interactions in electro-induced proton emission off nuclear targets.
A fully unfactorized relativistic formulation of Glauber
multiple-scattering theory is presented. The effect of truncating the
Glauber multiple-scattering series is discussed.  Relativistic effects
in the description of the final-state interactions are found not to
exceed the few percent level.  Also the frequently adopted approximation of
replacing the wave functions for the individual scattering nucleons by
some average density, is observed to have a minor impact on the
results.  Predictions for the separated $^4$He$(e,e'p)$ response
functions are given in quasi-elastic kinematics and a domain
corresponding with $1 \leq Q^2 \leq 2$~(GeV)$^2$.
   
\end{abstract}
\begin{keyword}
\PACS 25.30.Rw \sep 21.60.Cs \sep 24.10.Jv \sep 24.10.Ht 
\end{keyword}
\end{frontmatter}
%
% 25.30.Rw : Electroproduction reactions
% 21.60.Cs : Shell model 
% 24.10.Ht : Optical and diffraction models 
% 24.10.Jv : Relativistic model
%
%\pacs{25.30.Rw, 21.60.Cs, 24.10.Jv, 24.10.Ht}

\section{\label{intro} Introduction}

From early in the seventies till recent years, a systematic survey
with the exclusive $A(e,e'p)$ reaction on a whole range of target
nuclei revealed that the momentum distributions of bound protons in
nuclei are in line with the predictions of the nuclear mean-field
model.  The occupation probabilities for the single-particle levels,
on the other hand, turned out to be substantially smaller than what
could be expected within the context of a naive mean-field model
\cite{pandharipande97}.  This observation provided sound evidence for
the importance of short- and long-range correlations for the
properties of nuclei \cite{muether02}.  From 1990 onwards, the scope
of $A(e,e'p)$ reactions has been widened.  These days, rather than for
studying mean-field properties, electro-induced single-proton knockout
off nuclei is used as an experimental tool to learn for example about
relativistic effects in nuclei \cite{gao00} or to study fundamental
issues like possible medium modifications of protons and neutrons when
they are embedded in a dense hadronic medium like the nucleus
\cite{malov00,dieterich}.  Another issue is the question at
what distance scales quark and gluons become relevant degrees of
freedom for understanding the behavior of nuclei. Here, searches for
the onset of the color transparency phenomenon in $A(e,e'p)$ reactions
play a pivotal role \cite{Garrow02}.  In these studies one looks for
departures from predictions for proton transparencies from models
using standard nuclear-physics wave functions combined
with the best available tools for describing final-state interactions (FSI).

Traditionally, the results of exclusive $A(e,e'p)$ data-taking have
been interpreted with the aid of calculations within the framework of
the non-relativistic distorted wave impulse approximation (DWIA)
\cite{boffi93,Kelly96}.  In such an approach, the electromagnetic interaction
of the virtual photon with the target nucleus is assumed to occur
through the individual nucleons, an assumption which is known as the
impulse approximation (IA).  In the most simple DWIA versions, an
independent particle model (IPM) picture is adopted and the initial
and final A-nucleon wave functions are taken to be Slater
determinants.  The latter are composed of single-particle wave
functions which are solutions to a one-body Schr\"{o}dinger
equation.  Typically, in a DWIA approach the final proton scattering
state is computed as an eigenfunction of an optical potential,
containing a real and imaginary part.  Parameterizations for these
optical potentials are usually not gained from basic grounds, but
require empirical input from elastic $p + A \longrightarrow p + A$
measurements.  In the DWIA calculations, one adopts the philosophy
that potentials which parameterize FSI effects in elastic $p+A
\longrightarrow p+A$ reactions, can also be applied to model the
ejectile's distortions in $A+ e \longrightarrow A-1 + e' +p$.

A concerted research effort which started back in the late eighties
has resulted in the development of a number of relativistic DWIA
(RDWIA) models for computing $A(e,e'p)$ observables
\cite{picklesimer85,johansson96,yin92,udias93,Meucci2000}.  These
theoretical efforts very much followed the trend of developing
relativistic models for $p + A \longrightarrow p' + A$
processes as a potential improvement to the traditional
non-relativistic distorted-wave approaches.  Along the lines of the
DWIA approaches, the RDWIA frameworks are usually formulated within
the context of the independent-particle approximation with wave
functions of the Slater determinant form.  Relativistic bound-state
single-particle wave functions are customarily obtained within the
framework of the Hartree approximation to the $\sigma - \omega$ model
\cite{serot86}.  Scattering states by solving a time-independent Dirac
equation with relativistic optical potentials.  Systematic surveys
illustrated that in some $A(\vec{e},e'\vec{p})$ observables
relativistic effects are sizeable \cite{udias00,martinez02} and that
RDWIA-based calculations of $A(\vec{e},e'\vec{p})$ observables are at
least as successful as the more traditional non-relativistic DWIA
theories.

For proton lab momenta exceeding roughly 1 GeV, the use of optical
potentials for modeling FSI processes does not seem to be natural in
view of the highly inelastic character and diffractive nature of the
underlying elementary nucleon-nucleon scattering cross sections. In
this energy regime, an alternative description of FSI processes is
provided in terms of Glauber multiple-scattering theory
\cite{glauber70,Wallace75,Yennie71}.  This theory
essentially relies on the eikonal approximation and the assumption of
consecutive cumulative scattering of a fast proton on a composite
target containing A-1 ``frozen'' point scatterers.  Most $A(e,e'p)$
calculations based on the Glauber approach are non-relativistic and
factorized. A non-relativistic Glauber model for modeling
FSI effects in exclusive $^{4}$He$(e,e'p)$ reactions was recently
pointed out in Ref.~\cite{benhar00}.  A non-relativistic study of the
Glauber formalism in $d(e,e'p)n$ can for example be found in
Refs.~\cite{Claudio01} and \cite{jeschonnek99}.  For $(e,e'p)$
reactions off nuclei heavier than $^{12}$C and nuclear matter,
non-relativistic Glauber calculations have been reported in
Refs.~\cite{kohama00,frankfurt95,nikolaev95,Petraki2003}.

In this work we wish to present a relativistic formulation of Glauber
theory for calculating $A(e,e'p)$ observables.  The major assumptions
underlying our relativistic and unfactorized model bear a strong
resemblance with those adopted in the RDWIA models developed during
the last two decades.  One of the pivotal assumptions underlying
Glauber theory is the eikonal approximation.  Expressions of
Dirac-eikonal scattering amplitudes for elastic $p+A$ processes have
been derived in Ref.~\cite{amado83} and for $A(e,e'p)$ reactions in
Refs.~\cite{greenberg94,debruyne00}.  

The organization of this paper is as follows.  First, we briefly
sketch the $A(e,e'p)$ formalism in Sect.~\ref{sec:formalism}.  Next,
we introduce the relativistic formulation of Glauber
multiple-scattering theory in Sect.~\ref{sec:relglaub}.  Section
\ref{sec:results} is devoted to a presentation of results for the
Dirac-Glauber phase for the nuclei $^4$He, $^{12}$C, $^{56}$Fe and
$^{208}$Pb.  The Dirac-Glauber phase is a function which accounts for
all FSI effects when computing the $A(e,e'p)$ observables. The
contribution of single- and multiple-scatterings is estimated for the
target nuclei $^4$He, $^{12}$C and $^{208}$Pb.  Attention is paid to
the role of relativistic effects when computing the impact of
final-state interactions.  The validity of a frequently adopted
approximation, namely the replacement of the individual nucleon wave
functions by some average nuclear density, is investigated.  In
Sect.~\ref{sec:he4resp} we present predictions of our model for the
separated $(e,e'p)$ response functions for a $^4$He target nucleus.
Section \ref{sec:concl} summarizes our findings and states our
conclusions.     

\section{\label{sec:formalism} $A(e,e'p)$ Formalism}

We follow the conventions for the A$(\vec{e},e'\vec{p} \; )$B
kinematics and observables introduced by Donnelly and Raskin in Refs.~
\cite{donnelly86,raskin89}. The four-momenta of the incident and
scattered electron are denoted as $K^{\mu}_{e} (\epsilon,\vec{k})$ and
$K^{\mu}_{e'} (\epsilon',\vec{k'})$.  The electron momenta $\vec{k}$
and $\vec{k'}$ define the lepton scattering plane.  The four-momentum
transfer is given by $(\omega,\vec{q}\;) \equiv q^{\mu} = K^{\mu}_{e} -
K^{\mu}_{e'} = K_{A-1}^{\mu} + K_{p}^{\mu} - K_{A}^{\mu}$, where
$K_{A}^{\mu}(E_A,\vec{k}_A)$, $K_{A-1}^{\mu}(E_{A-1},\vec{k}_{A-1})$
and $K_{p}^{\mu}(E_p,\vec{k}_p)$ are the four-momenta of the target
nucleus, residual nucleus and the ejected proton.  The $z$-axis lies
along the momentum transfer $\vec{q}$ and the $y$-axis lies along
$\vec{k} \times \vec{k'}$. The hadron reaction plane is defined by
$\vec{k}_{p}$ and $\vec{q}$. The electron charge is denoted by $-e$, and we
adopt the standard convention $Q^{2} \equiv - q_{\mu} q^{\mu} \geq 0$
for the four-momentum transfer.

%The adopted normalization condition for the electron Dirac plane waves
%is
%\begin{equation}
%\bar{u}_e(K_{e},S)u_{e'}(K_{e'},S) = 1 \; .
%\end{equation}

In the laboratory frame, the differential cross section for $A + e
\longrightarrow (A-1) + e' + p$ processes can be computed from
\cite{donnelly86,raskin89,bjorken64} 
\begin{equation}
\frac{d^{5}\sigma}{d\epsilon'd\Omega_{e'}d\Omega_{p}} =
\frac{m_{e}^{2} M_{p} M_{A-1}}{(2\pi)^{5} M_{A}} \frac{k'k_{p}}{k}
f_{rec}^{-1} \: \overline{\sum_{if}} \left| {\mathcal M}_{fi} \right|^{2},
\end{equation}
where $f_{rec}$ is the hadronic recoil factor
\begin{equation}
f_{rec} = \frac{E_{A-1}}{E_{A}} \left| 1+\frac{E_{p}}{E_{A-1}}
\left(1-\frac{\vec{q} \cdot \vec{k}_{p}}{k_{p}^{2}} \right)\right| \; . 
\end{equation}
The squared invariant matrix element ${\mathcal M}_{fi}$ can be written as
\begin{equation}
\overline{\sum_{if}} \left| {\mathcal M}_{fi} \right|^{2} =
\frac{(4\pi\alpha)^{2}}{(Q^{2})^{2}} \, \eta_{e}
(K',S';K,S)_{\mu\nu}
W^{\mu\nu} ,
\end{equation}
where $\eta_{e} (K_{e'},S';K_{e},S)_{\mu\nu}$ is the electron tensor,
%defined
%by
%\begin{equation}
%\eta_{e} (K_{e'},S';K_{e},S)_{\mu\nu} \equiv  \overline{\sum_{if}} \left[
%\bar{u}_{e} (K_{e'},S') \gamma_{\mu} u_{e} (K_{e},S) \right]^{\star} \left[
%\bar{u}_{e} (K_{e'},S') \gamma_{\nu} u_{e} (K_{e},S) \right] \; ,
%\end{equation}
and the nuclear tensor $W^{\mu\nu} $ is given by
\begin{equation}
W^{\mu\nu} \equiv \overline{\sum_{if}} \left< J^{\mu} \right> ^{\dagger}
\left< J^{\nu} \right> \; ,
\end{equation}
with
\begin{equation}
\left< J^{\mu} \right>  = \left< A-1 (J_R \; M_R) , K_{p}(E_p, \vec{k}_p) m_s
\right| \widehat{J^{\mu}} \left| A(0^+,g.s.) \right> \; .
\end{equation}
Here, $\widehat{J ^{\mu}}$ is the electromagnetic current operator,
$\left| A(0^+,g.s.) \right>$ the ground state of the target even-even
nucleus and $\left| A-1 (J_R \; M_R) \right>$ the discrete state in
which the residual nucleus is left.

For $m_e / \epsilon \ll 1$ the contraction of the electron tensor
$\eta_{\mu\nu}$ with the nuclear one $W^{\mu\nu}$ results in an
expression of the form \cite{donnelly86}
\begin{equation}
4 m_{e}^{2} \eta_{e} (K_{e'},S';K_{e},S)_{\mu\nu} W^{\mu\nu}  = v_{0}
\sum_{K} v_{K} {\mathcal R}_{K} \; ,
\end{equation}
where the label $K$ takes on the values $L$, $T$, $TT$ and $TL$ and
refers to the longitudinal and transverse components of the virtual
photon polarization. The ${\mathcal R}_{K}$ are the nuclear response
functions and contain the nuclear structure and dynamics
information. Further, $v_{0} \equiv (\epsilon + \epsilon')^{2} -
q^{2}$ and the $v_{K}$ depend on the electron kinematics.  Combination
of the above results leads to the following final expression for the
$A(e,e'p)$ differential cross section \cite{donnelly86,raskin89}
\begin{eqnarray}
\frac{d^{5}\sigma} {d \epsilon' d \Omega_{e'} d \Omega_{p}}
 & =  & \frac{M _p M_{A-1}k_{p}}{8\pi^{3}M_{A}} f_{rec}^{-1} \sigma_{M}
\nonumber \\
& & \times \biggl[ v_{L}{\mathcal R}_{L}+v_{T}{\mathcal R}_{T}+v_{TT}{\mathcal
R}_{TT} \cos 2  \phi +v_{TL}{\mathcal R}_{TL} \cos \phi \biggr]  \; ,
\end{eqnarray}
where $\sigma_{M}$ is the Mott cross section
\begin{equation}
\sigma_{M} = \left( \frac{\alpha\cos\theta_{e}/2}{2\epsilon\sin^{2}\theta_{e}/2}\right)^{2} \; ,
\end{equation}
$\theta_{e}$ the angle between the incident and scattered electron and
$\phi$ the azimuthal angle of the plane define by $\vec{q}$ and
$\vec{k}_p$. The electron kinematics is contained in the kinematical
factors
\begin{eqnarray}
v_{L} & = & \left(\frac{Q^{2}}{q^{2}}\right)^{2} \; , \\ v_{T} & = &
\frac{1}{2} \left(\frac{Q^{2}}{q^{2}}\right) + \tan^{2}
\frac{\theta_{e}}{2} \; , \\ v_{TT} & = & -\frac{1}{2}
\left(\frac{Q^{2}}{q^{2}}\right) \; , \\ v_{TL} & = &
-\frac{1}{\sqrt{2}} \left(\frac{Q^{2}}{q^{2}}\right)
\sqrt{\left(\frac{Q^{2}}{q^{2}}\right)+ \tan^{2}
\frac{\theta_{e}}{2}} \; .
% \\ v_{T'} & = & \tan \frac{\theta_{e}}{2}
%\sqrt{\left(\frac{Q^{2}}{q^{2}}\right)+ \tan^{2}
%\frac{\theta_{e}}{2}} \; , \\ v_{TL'} & = & -\frac{1}{\sqrt{2}}
%\left(\frac{Q^{2}}{q^{2}}\right) \tan \frac{\theta_{e}}{2} \; .
\end{eqnarray}
The response functions are defined in the standard
fashion as
\begin{eqnarray}
{\mathcal R}_{L} & = & |\left< \rho(\vec{q}) \right>|^{2} \; , \\
{\mathcal R}_{T} & = & |\left< J(\vec{q};+1)\right> |^{2} +
| \left< J(\vec{q};-1) \right> |^{2} \; , \\
{\mathcal R}_{TT} \cos 2 \phi & = & 2 \, {\textstyle Re} \,
\biggl[ \left< J (\vec{q};+1) \right>^{\dagger} \left< J(\vec{q};-1) \right>
\biggr] \; , \\
{\mathcal R}_{TL} \cos \phi & = & -2 \, {\textstyle Re} \,
\Biggl[ \left< \rho  (\vec{q}) \right>^{\dagger}  \biggl(
\left< J(\vec{q};+1) \right> - \left< J(\vec{q};-1) \right>  \biggr) \Biggr] 
\; .
%{\mathcal R}_{T'} & = & \left| \left< J(\vec{q};+1) \right> \right |^{2} -
%\left|  \left< J(\vec{q};-1) \right> \right| ^ {2}\; , \\
%{\mathcal R}_{TL'} & = & -2 \, {\textstyle Re} \,
%\biggl[ \left< \rho (\vec{q}) \right>^{\dagger}  ( \left<
%J(\vec{q};+1) \right> + \left< J(\vec{q};-1) \right> )
%\biggr]
\end{eqnarray}
In the above expressions, $ \left< \rho (\; \vec{q} \; ) \right>
\equiv \left< J^{0}
(\; \vec{q} \; ) \right> $ denotes the Fourier transform of the transition
charge density $<f|\hat{\rho}(\vec{r})|i>$, while the Fourier
transform of the transition three-current density 
\begin{equation}
\left< f \left| \vec{J} (\; \vec{q} \; ) \right| i \right>  = \sum_{m
= 0, \pm 1} \left< f \left| J(\; \vec{q};m\; )
\vec{e}^{\; \dagger} _{m}  \right| i \right>  \; ,
\end{equation}
is expanded in terms of unit spherical vectors
$\vec{e} _ m$ 
\begin{equation}
\vec{e} _0 =  \vec{e}_{z} \; \; \; \; \; \; , 
\vec{e} _{\pm 1}  =  \mp \frac{1}{\sqrt{2}} (\vec{e}_{x}
\pm i \vec{e}_{y}) \; \; .
\end{equation}
Current conservation imposes the condition 
\begin{equation}
q_{\mu} J^{\mu} (\vec{q}) = \omega \rho (\vec{q}) - q J
(\vec{q};0) = 0 \; .
\end{equation}
All results presented in this paper are obtained in the Coulomb gauge
\begin{equation}
J^{\mu} = (\rho, J_x, J_y, \frac  {\omega} {q} \rho) \; ,
\end{equation}
and a relativistic current operator of the form
\begin{equation}
J^{\mu} = F_1 (Q^2) \gamma ^{\mu} + \frac { i \kappa} {2 M} F_2 (Q^2)
\sigma ^{\mu \nu} q _{\nu} \; , 
\end{equation}
where $F_1$ is the Dirac, $F_2$ the Pauli form factor and $\kappa$ the
anomalous magnetic moment.
 
\section{\label{sec:relglaub} Relativistic formulation of Glauber theory}
\subsection{\label{sec:glauscat} Dirac-eikonal approximation for
potential scattering}

We start our derivations by looking for solutions to the
time-independent Dirac equation for an ejectile with energy $E =
\sqrt{k^{2}_{p} + M^{2}}$ and spin state $\left| \frac {1} {2} m_s
\right> $ in the presence of spherical Lorentz scalar $V_{s}(r)$ and vector
potentials  $V_{v}(r)$
\begin{eqnarray}
[ \vec{\alpha} \cdot \vec{p}
+ \beta M + \beta V_{s} (r) + V_{v} (r) ] \Psi_{\vec{k}_{p},m_{s}}^{(+)} = E \Psi_{\vec{k}_{p},m_{s}}^{(+)}\; ,
\end{eqnarray}
where we have introduced the notation $\Psi_{\vec{k_p},m_{s}}^{(+)}$
for the unbound Dirac states. After some straightforward
manipulations, a Schr\"{o}dinger-like equation for the upper component
can be obtained
\begin{eqnarray}
\left[ -\frac{\hbar^{2}\nabla^{2}}{2M} + V_{c} + V_{so} (\vec{\sigma} \cdot \vec{L} -
i \vec{r} \cdot \vec{p} ) \right]  u ^{(+)} (\vec{k}_p,m_s) =
\frac{k^{2}_{p}}{2M} u ^{(+)} (\vec{k}_p,m_s) \; ,
\label{eq:uppereq}
\end{eqnarray}
where the central and spin orbit potentials $V_{c}$ and $V_{so}$ are
defined as
\begin{eqnarray}
\label{eq:vsandvso}
V_{c} (r) & = & V_{s} (r) + \frac{E}{M} V_{v} (r) + \frac{V_{s} ^{2}
(r) - V_{v}^{2} (r)}{2M} \; , \nonumber \\ V_{so} (r) & = &
\frac{1}{2M[E+M+V_{s}(r)-V_{v}(r)]} \frac{1}{r} \frac{d}{dr}
[V_{v}(r)-V_{s}(r)] \; .
\end{eqnarray}

Since the lower component $w ^{(+)} (\vec{k}_p,m_s)$ is related to the upper one through
\begin{eqnarray}
w ^{(+)} (\vec{k}_p,m_s) = \frac{1}{E+M+V_{s}-V_{v}} \vec{\sigma} \cdot
\vec{p} \, u ^{(+)} (\vec{k}_p,m_s) \; ,
\end{eqnarray}
the solutions to the Eq.~(\ref{eq:uppereq}) determine the scattering
state.  In RDWIA approaches, a Dirac equation of the type
(\ref{eq:uppereq}) is solved numerically for Dirac optical potentials
$V_{s}(r)$ and $V_{v}(r)$ derived from global fits to elastic proton
scattering data ~\cite{cooper93}.  Not only are global
parameterizations of Dirac optical potentials usually restricted to
proton kinetic energies $T_{p} \leq 1$~GeV, calculations based on
exact solutions of the Dirac equation frequently become impractical at
higher energies.  This is particularly the case for approaches relying
on partial-wave expansions.  At higher proton kinetic energies it
appears more convenient to solve the Dirac equation (\ref{eq:uppereq})
in the eikonal approximation
\cite{benhar00,amado83,greenberg94,ito97}.  In intermediate-energy
elastic $p-^{40}$Ca scattering ($T_p \approx$ 500~MeV) the eikonal
approximation was shown to successfully reproduce the exact Dirac
partial-wave result. Bianconi and Radici showed that for ejectile
momenta exceeding 1~GeV, the eikonal approximation almost reproduced
the $^{12}$C$(e,e'p)$ differential cross sections obtained through
performing a partial-wave expansion of the ``exact'' scattering wave
function \cite{bianconi95,bianconi96}.

As shown for example in Refs. \cite{amado83} and \cite{debruyne00}, in
the eikonal limit the scattering wave function takes on the form
\begin{eqnarray}
\psi_{\vec{k}_{p},m_{s}}^{(+)} = \sqrt { \frac {E+M} {2M} } 
\left[
\begin{array}{c}
1 \\
\frac{1}{E+M+V_{s}-V_{v}} \vec{\sigma} \cdot \vec{p}
\end{array}
\right]
e^{i \vec{k}_{p} \cdot \vec{r}} e^{i S(\vec{r})}
\chi_{\frac{1}{2}m_{s}} \; ,
\label{eq:thescatteringwave}
\end{eqnarray}
where the eikonal phase reads ($\vec{r} \equiv (\vec{b},z)$)
\begin{eqnarray}
i S(\vec{b},z) = - i \frac{M}{K} \int_{-\infty}^{z} dz' \,
\biggl[ V_{c} (\vec{b},z') + V_{so} (\vec{b},z') [ \vec{\sigma} \cdot
(\vec{b} \times \vec{K} )- i Kz'] \biggr] .
\label{eq:eikonalphase}
\end{eqnarray}
In modeling $A(e,e'p)$ processes, the average momentum $\vec{K}$
occurring in this expression is defined as
\begin{equation}
\vec{K} = \frac {\vec{k} _p + \vec{q} } {2} \; ,
\end{equation}
where $\vec{q}$ is the three-momentum transfer induced by the virtual
photon.  The scattering wave function from
Eq.~(\ref{eq:thescatteringwave}) differs from the plane wave solution
in two respects.  First, the lower component exhibits the dynamical
enhancement due to the combination of the scalar and vector
potentials.  Second, the eikonal phase $e^{i S(\vec{r})}$ accounts for
the interactions that the struck nucleon undergoes in its way out of
the target nucleus.  The eikonal approximation is a valid one, if the
magnitude of the three-momentum transfer $\mid \vec{q} \mid $ is
sufficiently large in comparison with the projected initial (or,
missing) momentum of the ejectile (or, $ q \gg p_m = \mid \vec{k}_p -
\vec{q} \mid $).  The eikonal phase of Eq.~(\ref{eq:eikonalphase})
reflects the accumulated effect of all interactions which the ejectile
undergoes in its way out of the nucleus.  All these effects are
parametrized in terms of one-body optical potentials and the link with
the elementary proton-proton and proton-neutron scattering processes
appears lost.  In Glauber theory this link with the elementary
processes is reestablished.

\subsection{\label{sec:glaupN} Proton-nucleon scattering}

To start our derivations of a relativistic version of Glauber
multiple-scattering, we first consider a nucleon-nucleon scattering
process and assume that it is governed by a local Lorentz and vector
potential $V_s(r)$ and $V_v(r)$.  In the eikonal approximation, the
scattering amplitude corresponding with this process reads
\cite{amado83}
\begin{eqnarray}
F_{m_{s}m_{s'}} (\vec{k}_{i},\vec{k}_{f}) = - \frac{M}{2\pi}
\left< \psi_{\vec{k}_{f},m_{s'}}^{(+)}   \right| (\beta V_{s} + V_{v}) \left|
  \Phi_{\vec{k}_{i},m_{s}} \right> \; ,
\end{eqnarray}
with a relativistic scattered wave $\psi_{\vec{k}_{f},m_{s'}}^{(+)}$
as determined in the eikonal approximation (Eq.~(\ref{eq:thescatteringwave})) 
and the free Dirac solution
\begin{eqnarray}
\Phi_{\vec{k}_{i},m_{s}} = \sqrt{\frac{E+M}{2M}}
\left[
\begin{array}{c}
1 \\
\frac{1}{E+M} \vec{\sigma} \cdot \vec{p}
\end{array}
\right]
e^{i \vec{k}_{i} \cdot \vec{r}}
\chi_{\frac{1}{2}m_{s}} \; .
\end{eqnarray}
After some algebraic manipulations one finds \cite{amado83}
\begin{eqnarray}
F_{m_{s},m_{s'}} (\vec{k}_{i},\vec{k}_{f}) = -i K \left< m_{s'} \left|
\int \frac{d\vec{b}}{2\pi} e^{-i
\left( \vec{k}_f - \vec{k}_i \right) \cdot\vec{b}}
(e^{i\chi(\vec{b})}-1) \right| m_{s} \right> \; ,
\label{eq:Fss}
\end{eqnarray}
where the phase shift function is given by
\begin{eqnarray}
\chi(\vec{b}) =  \frac{M}{K} \int_{-\infty}^{+\infty} dz \, [
V_{c} (\vec{b},z) + V_{so} (\vec{b},z) [ \vec{\sigma} \cdot (\vec{b}
\times \vec{K} )]] \; .
\end{eqnarray}
In conventional Glauber theory the phase shift function $\chi(\vec{b})$ is
not calculated on the basis of knowledge about the radial dependence
and magnitude of the potentials $V_c(r)$ and $V_{so}(r)$, but is directly
extracted from proton-proton and proton-neutron scattering data.
This requires some extra manipulations which will be exposed below.  

\begin{figure}
\begin{center}
\resizebox{0.75\textwidth}{!}{\includegraphics{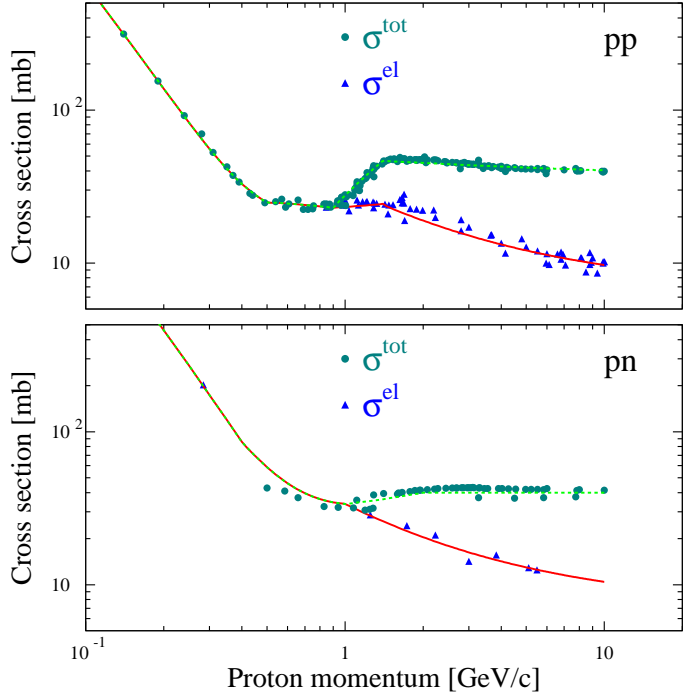}}
\end{center}
\caption{\label{fig:partotcross} Total and elastic cross sections for
proton-proton and proton-neutron scattering as a function of the
proton lab momentum.  The data are from Ref.~\cite{pdg}. The solid
(dashed) curve displays our global fit to the elastic (total) cross
section.}
\end{figure}

Using rotational invariance and parity conservation, a $pN$
scattering process where at most one polarization of the colliding
particles is determined, can be analyzed with a scattering amplitude
of the form \cite{alkhazov78}
\begin{eqnarray}
f_{pN} (\Delta) = f_{pN}^{c} (\Delta) + f_{pN}^{s} (\Delta)
\vec{\sigma} \cdot \left( \vec{k_i} \times \vec{k_f} \right)  \; ,
\end{eqnarray}
where $f_{pN}^{c}$ and $f_{pN}^{s}$ are the central and spin-dependent
amplitudes, $\vec{\sigma}$ is the spin-operator corresponding with the
incident proton, and $\vec{\Delta} = \vec{k}_f - \vec{k}_i $ is the
transferred momentum.  The small angle elastic scattering of protons
with $k_p > 1$~GeV is dominated by the central, spin-independent
amplitude $f_{pN}^{c}$ \cite{alkhazov78}.  Given the diffractive
nature of $pN$ collisions at these energies, the central amplitudes
are usually parameterized in a functional form of the type
\begin{eqnarray}
f_{pN}^{c} (\Delta) = \frac{k_{f} \sigma^{tot}_{pN}}{4\pi} (\epsilon_{pN} + i)
\exp\left( - \frac{\Delta^{2} \beta_{pN}^{2}}{2} \right).
\label{eq:fparam}
\end{eqnarray}
The parameters in Eq.~(\ref{eq:fparam}) can be determined from fitting
the results of proton-nucleon scattering experiments. A selection of
the measured elastic $(\sigma ^{el})$ and total $(\sigma ^{tot})$
cross sections for proton-proton and proton-neutron scattering are
shown in Fig.~\ref{fig:partotcross}.  The above form for the
scattering amplitude $f_{pN}^{c}$ corresponds with a differential
cross section which reads at forward angles ($t \equiv \left( k_f
^{\mu} - k_i ^{\mu} \right) ^2$)
\begin{equation}
\frac {d \sigma _{pN} ^{el}}  {d t}  \approx \left. \frac {d \sigma
_{pN} ^{el}} {d
t} \right| _{t=0} \exp ^{ -  \beta _{pN} ^2 \mid t \mid } \; .  
\end{equation} 

\subsection{\label{sec:glaurelA} Relativized Glauber model for $A(e,e'p)$}
In the relativized Glauber multiple scattering framework, the
antisymmetrized $A$-body wave function in the final state reads
\begin{eqnarray}
\Psi ^{\vec{k}_{p},m_{s}}_{A} \left( \vec{r}, \vec{r}_2, \vec{r}_3,
\ldots \vec{r}_A \right) & \sim & {\widehat{\mathcal{A}}} \Biggl[
{\widehat{\mathcal{S}} \left( \vec{r}, \vec{r}_2, \vec{r}_3, \ldots
\vec{r}_A \right) } \left[
\begin{array}{c}
1 \\
\frac{1}{E+M} \vec{\sigma} \cdot \vec{p} 
\end{array} 
\right] e^{i \vec{k}_p \cdot \vec{r}} \chi_{\frac{1}{2}m_{s}}
\nonumber \\ & & \times \Psi _{A-1} ^{J_R \; M_R} \left( \vec{r}_2,
\vec{r}_3, \ldots \vec{r}_A \right) \Biggr] \; ,
\label{eq:glauber}
\end{eqnarray}
where $\Psi _{A-1} ^{J_R \; M_R}$ is the wave function characterizing
the state in which the $A-1$ nucleus is created and
${\widehat{\mathcal{A}}}$ is the antisymmetrization operator.  In the
above expression, the subsequent elastic or ``mildly inelastic''
collisions of the struck proton with the ``frozen'' spectator nucleons, are
implemented through the introduction of the following operator
\begin{equation}
{\widehat{\mathcal{S}} \left( \vec{r}, \vec{r}_2,
\vec{r}_3, \ldots \vec{r}_A \right) }
\equiv \prod _ {j=2} ^{A} \left[ 1 - \Gamma \left( \vec{b} - \vec{b_j}
\right) \theta \left( z_j - z \right) \right] \; ,
\label{eq:glauoperator}
\end{equation}
where $\vec{r} \left( \vec{b},z \right)$ denotes the position of the
struck particle and $\left( \vec{r}_2, \vec{r} _3, \ldots , \vec{r}_A
\right) $ those of the (frozen) spectator protons and neutrons in the
target. The step function $\theta \left( z_j -z \right) $ guarantees
that the struck proton can only interact with the spectator protons
and neutrons which it finds in its forward propagation path.  Further,
the profile function for $pN$ scattering is defined as
\begin{eqnarray*}
\Gamma (\vec{b}) \equiv 1 - e ^{i \chi (\vec{b}) } = \frac{{\sigma^{tot}_{pN}}
(1-i {\epsilon_{pN}})}
{4\pi \beta_{pN}^{2}} \: exp \left(
-\frac{b^{2}}{2\beta_{pN}^{2}}  \right) \; ,
\label{eq:profilefunction}
\end{eqnarray*}
where the last equality can be derived from the
expression~(\ref{eq:fparam}).  Two major assumptions underly the
derivation of the operator of Eq.~(\ref{eq:glauoperator}) : the
eikonal approximation for the continuum wave function of the struck
proton and the frozen approximation for the positions of the
spectators.  Indeed, one assumes that the ejectile passes through the
nucleus in a very short time so that variations in the positions of
the residual nucleons can be ignored. The operator of
Eq.~(\ref{eq:glauoperator}) represents the accumulated effect of the
phase shifts contributed by each of the target scatterers as the
ejectile progresses through the nucleus.  The property of so-called
phase-shift additivity is a direct consequence of the adopted
one-dimensional nature of the relative motion, together with the
neglect of three- and more-body forces, recoil effects and
longitudinal momentum transfer.

The Dirac-Glauber $A(e,e'p)$ transition amplitude can be written as
\begin{eqnarray}
\left< J^{\mu} \right> &= & \int d \vec{r} \int d \vec{r}_2 \ldots \int d
\vec{r}_A \nonumber \\
& & \times 
\left( \Psi ^{\vec{k}_p , m_{s}}_{A} \left( \vec{r}, \vec{r}_2, \vec{r}_3, \ldots
\vec{r}_A \right) \right) ^ {\dagger} \gamma ^{0}
J^{\mu} (r) e ^{ i \vec{q} \cdot \vec{r} }
 \Psi ^{gs}_{A} \left( \vec{r}, \vec{r}_2, \vec{r}_3, \ldots
\vec{r}_A \right) \; ,
\label{eq:diracglauber}
\end{eqnarray}
where for convenience only the spatial coordinates are explicitly
written.  The difficulty of the evaluation of this matrix element
stems from the fact that the multiple-scattering operator
${\widehat{\mathcal{S}}}$ is a genuine $A$-body operator.  One popular
approximation in Glauber inspired $A(e,e'p)$ calculations, is
expanding the A-body operator ${\widehat{\mathcal{S}} \left( \vec{r},
\vec{r}_2, \vec{r}_3, \ldots \vec{r}_A \right) }$
\begin{eqnarray}
& & {\widehat{\mathcal{S}}   \left( \vec{r}, \vec{r}_2,
\vec{r}_3, \ldots \vec{r}_A \right) } =    
1 - \sum_{j=2}^A\theta(z_j-z)\Gamma(\vec{b} - \vec{b}_j) 
\nonumber \\
& + & \sum _
{j\neq k} ^{A} \theta(z_j-z) \Gamma(\vec{b} - \vec{b}_j)
\theta(z_k-z) \Gamma(\vec{b} -\vec{b}_k) \nonumber \\
& - & \sum_
{j\neq k \neq l =2 } ^{A} \theta(z_j-z) \Gamma(\vec{b} - \vec{b}_j)
\theta(z_k-z) \Gamma(\vec{b} - \vec{b}_k) \theta(z_l-z) \Gamma(\vec{b}
- \vec{b}_l)  \nonumber \\ & + &  \ldots \; ,
\label{eq:glaubtrunc}
\end{eqnarray}
and truncating it at some order in $\Gamma$.  Formally, this expansion
bears a strong resemblance with the Mayer-Ursell expansion used in
modeling the correlation effects in the theory of real gases and
liquids. In the above expression, the unity operator (first term)
reflects the situation whereby the ejectile is not subject to
scatterings in its way out of the nucleus. The second term, which is
linear in the profile function, accounts for processes whereby the
struck nucleon scatters on one single spectator nucleon before turning
asymptotically free (single-scattering process). Higher-order terms in
the expansion refer to processes whereby the ejected proton
subsequently scatters with two, three, $\ldots$, A-1 spectator
nucleons.  Evaluating the different terms in the above expansion
allows one to estimate the effect of single-, double, triple-,
$\ldots$ scatterings. For heavier nuclei, truncating the expansion at
first order in $\Gamma$ appears as a rather questionable procedure.
In Section \ref{sec:results} results obtained with the truncated
(Eq.~(\ref{eq:glaubtrunc})) and complete (Eq.~(\ref{eq:glauoperator}))
Dirac-Glauber multiple-scattering operator will be compared.

In evaluating the Dirac-Glauber transition amplitude of
Eq.~(\ref{eq:diracglauber}) we have introduced a minimal amount of
approximations.  In line with the assumptions underlying the RDWIA
approaches we adopt a mean-field approximation for the nuclear wave
functions.  For the sake of brevity of the notations, in the
forthcoming derivations we consider the case $A=3$.  A generalization
to arbitrary mass number $A$ is rather straightforward.  Further, we
will assume that the nuclear current $J^{\mu}$ is a one-body
operator. As both the initial and final wave functions are fully
antisymmetrized, one can choose the operator $J^{\mu}$ to act on one
particular coordinate and write without any loss of generality
$\left( \vec{r} \; ' \equiv (\vec{b}'(x',y'),z') \right)$
\begin{eqnarray}
\left< J^{\mu} \right> & = & A \frac {1} {A!} \int d \vec{r} \; ' \int
d \vec{r}_2 \hspace{0.1mm} ' \int d
\vec{r} _3 \hspace{0.1mm} ' 
\sum _ { k,l,m \in \left\{ k_{p} m_{s}, \alpha_2, \alpha _3 \right\} } 
\; \; \sum _ { n,o,p \in \left\{ \alpha_1, \alpha_2, \alpha _3 \right\} }
\epsilon _{klm}^{*} \epsilon _{nop} \nonumber \\ & & 
\times \phi _ {k} ^{\dagger} \left( \vec{r} \; '   \right)
\phi _ {l} ^{\dagger} \left( \vec{r}_2 \hspace{0.1mm} ' \right)
\phi _ {m} ^{\dagger} \left( \vec{r}_3 \hspace{0.1mm} ' \right)
e ^{ i \vec{q} \cdot \vec{r} \; ' } \gamma _0
\nonumber \\ & & \times J^{\mu}(\vec{r} \; ') 
 \left[ 1 - \theta \left( z'_2 - z' \right) \Gamma \left( \vec{b}' _2 -
\vec{b}' \right) \right] ^{\dagger}
\left[ 1 - \theta \left( z'_3 - z' \right) \Gamma \left( \vec{b}' _3 -
\vec{b}' \right) \right] ^{\dagger}
\nonumber \\ & & \times 
\phi _ {n} \left( \vec{r} \; '     \right)
\phi _ {o} \left( \vec{r}_2 \hspace{0.1mm} '   \right)
\phi _ {p} \left( \vec{r}_3 \hspace{0.1mm} '  \right) \; ,
\label{eq:startder}
\end{eqnarray}  
with $\epsilon _{ijk}$ the Levi-Civita symbol, and where we
have introduced a frame $(x',y',z')$ defined by the following unit vectors
\begin{equation}
\hat{z}' =  \frac {\vec{k}_p} {\mid \vec{k}_p \mid} \; \; \; \; , \;
\;  
\hat{y}' = \frac {\vec{k}_p \times \vec{q}} 
{ \mid \vec{k}_p 
\times \vec{q} \mid } \; \; \; \; , \; \; 
\hat{x}' = \hat{z}' \times \hat{y}' \; .
\label{eq:eikonalframe}
\end{equation}
The plane $(x',z')$ coincides with what is usually known as the
hadron reaction plane in $A(e,e'p)$ reactions.

Adopting a mean-field picture, the initial $A$-nucleon wave function
is of the form
\begin{eqnarray}
\Psi ^{gs}_{A} \left( \vec{r}_1, \vec{r}_2, \vec{r}_3 \right)
= \frac {1} {\sqrt{A!}} \left|
\begin{array}{ccc}
  \; \; \phi _{\alpha_{1}} \left( \vec{r}_1  \right) \; \;  
& \; \; \phi _{\alpha_{2}} \left( \vec{r}_1  \right) \; \;  
& \; \; \phi _{\alpha_{3}} \left( \vec{r}_1  \right) \; \;  \\
  \; \; \phi _{\alpha_{1}} \left( \vec{r}_2  \right) \; \;  
& \; \; \phi _{\alpha_{2}} \left( \vec{r}_2  \right) \; \;  
& \; \; \phi _{\alpha_{3}} \left( \vec{r}_2  \right) \; \;  \\
  \; \; \phi _{\alpha_{1}} \left( \vec{r}_3  \right) \; \;  
& \; \; \phi _{\alpha_{2}} \left( \vec{r}_3  \right) \; \;  
& \; \; \phi _{\alpha_{3}} \left( \vec{r}_3  \right) \; \;  
\end{array} \right| \; .
\label{eq:groundstate}
\end{eqnarray} 
For spherically symmetric potentials, the solutions 
$\phi _{\alpha} \left( \vec{r}  \right)$ to a
single-particle Dirac equation entering this Slater determinant have
the form \cite{Walecka2001}
\begin{equation}
\phi_{\alpha} \left( \vec{r} , \vec{\sigma} \right) \equiv \phi_{n
\kappa m } ( \vec{r} , \vec{\sigma} ) = \left[
\begin{array}{c}
i \frac {G_{n \kappa}  ( r )} {r}\; {\mathcal Y}_{\kappa m} (\Omega, \vec{\sigma}) \\
- \frac {F_{n \kappa} ( r )} {r } \; {\mathcal Y}_{- \kappa m} (\Omega, \vec{\sigma}) 
\end{array} 
\right] \; ,
\label{eq:diracsingle}
\end{equation}
where $n$ denotes the principal, $\kappa$ and $m$ the generalized
angular momentum quantum numbers. The ${\mathcal
Y}_{\pm \kappa m}$ are the spin spherical harmonics and
determine the angular and spin parts of the wave function,
\begin{eqnarray}
{\mathcal Y}_{\kappa m} (\Omega, \vec{\sigma}) = \sum_{m_{l}m_{s}}
\left<lm_{l}\frac{1}{2}m_{s} \mid jm \right> Y_{lm_{l}} (\Omega)  \chi_{\frac{1}{2}
m_{s}} (\vec{\sigma}) \; , \nonumber \\ 
j = |\kappa| - \frac{1}{2} \; , \hspace{0.8cm} l = \left\{
\begin{array}{ll}
\kappa, & \kappa > 0 \\
-(\kappa+1), & \kappa < 0 \; .
\end{array}
\right.
\end{eqnarray}
The final $A$-body wave function reads 
\begin{eqnarray}
\Psi ^{\vec{k}_{p},m_{s}}_{A}  \left( \vec{r}_1 , \vec{r}_2 , \vec{r}_3
\right) & = & 
\frac {1} {\sqrt{A!}} \left| 
\begin{array}{ccc}
\; \; {\widehat{\mathcal{S}}} \left( \vec{r}_1, \vec{r}_2,
\vec{r}_3 \right) \; \;  \phi _{k_{p} m_{s}} \left( \vec{r}_1 \right) \; \;  & \; \; \phi _
{\alpha_{2}} \left( \vec{r}_1  \right) \; \;  & \; \; \phi _
{\alpha_{3}} \left( \vec{r}_1 \right) \; \;  \\
\; \; {\widehat{\mathcal{S}}} \left( \vec{r}_2, \vec{r}_1,
\vec{r}_3 \right) \; \;  \phi _{k_{p} m_{s}} \left( \vec{r}_2 \right) \; \;  & \; \; \phi _
{\alpha_{2}} \left( \vec{r}_2  \right) \; \;  & \; \; \phi _
{\alpha_{3}} \left( \vec{r}_2 \right) \; \;  \\
\; \; {\widehat{\mathcal{S}}} \left( \vec{r}_3, \vec{r}_1,
\vec{r}_2 \right) \; \;  \phi _{k_{p} m_{s}} \left( \vec{r}_3 \right) \; \;  & \; \; \phi _
{\alpha_{2}} \left( \vec{r}_3  \right) \; \;  & \; \; \phi _
{\alpha_{3}} \left( \vec{r}_3 \right) \; \;  
\end{array} \right| \; . \nonumber \\
\label{eq:finalstate}
\end{eqnarray}
Relative to the target nucleus ground state written in
Eq.~(\ref{eq:groundstate}), the wave function of
Eq.~(\ref{eq:finalstate}) refers to the situation whereby the struck
proton resides in a state ``$\alpha _1$'', leaving the residual $A-1$
nucleus as a hole state in that particular single-particle level.

Assuming that the profile function $\Gamma$ does not contain
spin-dependent terms, one can safely assume that for elastic and
mildly inelastic scatterings
\begin{eqnarray}
\int & & d \vec{r}_{1} \hspace{0.1mm} ' \int d \vec{r}_{2} \hspace{0.1mm} ' 
\phi _ {k} ^{\dagger} \left( \vec{r}_{1} \hspace{0.1mm} '   \right)
\phi _ {l} ^{\dagger} \left( \vec{r}_{2} \hspace{0.1mm} '   \right)
J^{\mu}(\vec{r}_1 \hspace{0.1mm} ') 
\nonumber \\ 
& &
\times \left[ 1 - \theta \left( z_2 ' - z_1 ' \right) \Gamma \left(
\vec{b}_2  ' -
\vec{b}_1  ' \right) \right] ^{\dagger}
\phi _ {n} \left( \vec{r}_{1} \hspace{0.1mm} '   \right)
\phi _ {o} \left( \vec{r}_{2} \hspace{0.1mm} '   \right) 
\nonumber \\ 
& & \approx
\delta _{lo} \int d \vec{r}_1 \hspace{0.1mm} ' \int d \vec{r}_{2} \hspace{0.1mm}  ' 
\phi _ {k} ^{\dagger} \left( \vec{r}_{1} \hspace{0.1mm} '   \right)
J^{\mu}(\vec{r} _1 \hspace{0.1mm} ') 
\nonumber \\
& & \times \left[ 1 - \theta \left( z_2 ' - z_1 ' \right) \Gamma
\left( \vec{b}_2 ' -
\vec{b}_1 ' \right) \right] ^{\dagger}
\phi _ {n} \left( \vec{r}_{1}  \hspace{0.1mm} '   \right)
\left| \phi _ {o} \left( \vec{r}_{2} \hspace{0.1mm} '   \right) \right|^2 \; .
\end{eqnarray}
Inserting this expression in Eq.~(\ref{eq:startder}) one obtains
\begin{eqnarray}
\left< J^{\mu} \right> & = & A \frac {1} {A!} \int d \vec{r}
\; ' \int d \vec{r}_{2} \hspace{0.1mm} ' \int d
\vec{r} _3 \hspace{0.1mm} ' 
\sum _ { l,m \in \left\{ \alpha_2, \alpha _3 \right\} } 
\epsilon _{(k_{p}m_{s})lm}^{*} \epsilon _{\alpha _{1} l m} \nonumber 
\\ & & 
\times \phi _ {k_{p} m_{s}} ^{\dagger} \left( \vec{r} \;  '   \right)
\left| \phi _ {l} \left( \vec{r} _{2} \hspace{0.1mm}  ' \right) \right|^2
\left| \phi _ {m} \left( \vec{r} _{3} \hspace{0.1mm}  ' \right) \right|^2
e ^{ i \vec{q} \cdot \vec{r} \;  ' } \gamma _0
J^{\mu}(\vec{r} \;  ') 
\phi _ {\alpha_{1}} \left( \vec{r} \;  '     \right)
\nonumber \\ & & \times  
 \left[ 1 - \theta \left( z_2 ' - z ' \right) \Gamma \left( \vec{b}_2 ' -
\vec{b}' \right) \right]^{\dagger}
\left[ 1 - \theta \left( z_3 ' - z ' \right) \Gamma \left( \vec{b}_3 ' -
\vec{b}' \right) \right]^{\dagger}
 \; .
\end{eqnarray}
This leads to our final result for the Dirac-Glauber $A(e,e'p)$ transition
amplitude
\begin{equation}
\left< J^{\mu} \right>  =   \int d \vec{r} \phi _ {k_{p}m_{s}} ^{\dagger}
\left( \vec{r} \right)
\mathcal{G}^{\dagger}(\vec{b},z)  \gamma ^{0} J^{\mu}(\vec{r})
e ^{ i \vec{q} \cdot \vec{r} } 
\phi _ {\alpha_{1}} \left( \vec{r}
\right) \; ,
\label{eq:glauampfinal}
\end{equation}
where the Dirac-Glauber phase $\mathcal{G}(\vec{b},z)$ is defined in the
following fashion
\begin{equation}
\mathcal{G}(\vec{b},z)=
\prod _{\alpha_{occ} \ne \alpha} \left[
1 - \int d \vec{r} \hspace{0.1mm} '
\left| \phi _ {\alpha_{occ}} \left( \vec{r} \hspace{0.1mm} '     \right) \right|^2 
 \theta \left( z' - z \right) \Gamma \left( \vec{b}' -
\vec{b} \right)  \right] \; .
\label{eq:glauberphase}
\end{equation}
In this expression, the product extends over all occupied
single-particle states, except for the one from which the nucleon is
ejected.  The RPWIA approximation would correspond with putting
$\mathcal{G}=1$ in the expression for the matrix element of
Eq.~(\ref{eq:glauampfinal}).  

The numerical evaluation of the Glauber phase
$\mathcal{G}(\vec{b},z)$ is rather challenging if no additional
approximations are introduced. A Monte Carlo integration method was
suggested in Ref~\cite{varga2002}. In our numerical calculations we
did not introduce any further approximations and found it most
appropriate to evaluate the scattering amplitudes and Glauber phases
in the frame defined by the unit vectors of
Eq.~(\ref{eq:eikonalframe}).  Inserting the expression for the Dirac
single-particle wave functions $\phi _{\alpha}$ of
Eq.~(\ref{eq:diracsingle}) in the Eq.~(\ref{eq:glauberphase}) for
the Glauber phase, one gets ($ d \vec{r} \hspace{0.3mm} ' \equiv dz'b' db' d
\phi_{b'}$)
\begin{eqnarray}
\mathcal{G} & & (\vec{b},z) =  \prod _{\alpha_{occ} \ne \alpha}  
\Biggl\{ 1 - 
\frac{\sigma^{tot}_{pN}( 1 - i \epsilon _{pN})} {4\pi \beta _{pN}^2} 
\int _{0} ^{\infty} b'db' \int _{0}
^{\infty} dz' \theta (z' -z) 
\nonumber \\
& &  \times \Biggl( \biggl[ \frac{G_{n \kappa} \left( r'(b',z') \right)}
{r'(b',z')}
\mathcal{Y} _{\kappa m} (\Omega(b',z') , \vec{\sigma})
\biggr] ^2 \nonumber \\
& & + \biggl[ \frac{ F_{n \kappa} \left(
r'(b',z') \right) }
{r'(b',z')}
\mathcal{Y} _{\kappa m} (\Omega (b',z'), \vec{\sigma}) \biggr]  ^2 \Biggr)  
\nonumber \\ & & \times
\exp \left[ -\frac{(b - b')^2 }{2 \beta _{pN} ^2} \right] 
\int _{0} ^{2 \pi} d \phi_{b'}
\exp \biggl[ \frac{-b b'}{\beta_{pN}^2}2 {\sin}^2 \left( \frac {\phi_{b} -\phi_{
b'}} {2} \right) \biggr]
\Biggr\} \; .
\label{eq:glauberphase1}
\end{eqnarray}
Standard numerical integration techniques were adopted to evaluate the
integrals occurring in this equation.  It is important to remark that
cylindrical symmetry about the $z'$ axis makes the above expression to
be independent of $\phi _b$.  As a result the relativistic Glauber
depends on only two independent variables $(b,z)$.  In the above
expression~(\ref{eq:glauberphase1}) each of the frozen spectator
nucleons is identified by its quantum numbers $(n, \kappa,m)$ and its
corresponding Dirac wave function $\phi _{n \kappa m} (\vec{r},
\vec{\sigma})$. In most Glauber-based calculations, an additional
averaging over the positions of the spectator nucleons is introduced.
This procedure amounts to replacing in Eq.~(\ref{eq:glauberphase1})
the characteristic spatial distributions of each of the spectator
nucleons described by the functions $F_{n \kappa}(r)$ and $G _{n
\kappa} (r)$ by an average density distribution for the target nucleus
\begin{eqnarray}
\mathcal{G} (b,z) & =&  
\Biggl\{ 1 - 
\frac{\sigma^{tot}_{pN}( 1 - i \epsilon _{pN})} {4\pi \beta _{pN}^2} 
\int _{0} ^{\infty} b'db'
T_B (b',z)
 \exp \left[ -\frac{(b - b')^2 }{2 \beta _{pN} ^2} \right] 
\nonumber \\
& & \times
\int _{0} ^{2 \pi} d \phi_{b'}
\exp \biggl[ \frac{-b b'}{\beta_{pN}^2}2 {\sin}^2 \left( \frac {\phi_{b} -\phi_{
b'}} {2} \right) \biggr]
\Biggr\}^{A-1} \; .
\end{eqnarray}
The function $T_B(b',z)$ which was introduced in the above expressions
is known as the ``thickness function'' and reads 
\begin{equation}
T_B (b',z) = \frac {1} {A} \int _0 ^{+ \infty} d z' \theta \left( z' -
z \right) \rho_B (r'(b',z'))   \; ,
\end{equation}
where the relativistic radial baryon density $\rho _B (r) $ is defined in the
standard fashion
\begin{eqnarray}
\rho _B ( r ) \equiv \left< \overline{\Psi _A ^{gs} } \gamma _0 \Psi _A
^{gs} \right> & = & \sum _{\alpha _{occ}} \int d \vec{\sigma} d \Omega \left( \phi_{\alpha _{occ}}
(\vec{r}, \vec{\sigma})
\right) ^ {\dagger}  
\left( \phi_{\alpha _{occ}}
(\vec{r}, \vec{\sigma})
\right) \; \nonumber \\
& = & \sum _{n \kappa} \frac { (2 j +1 ) } {4 \pi r^2} \biggl[ \left|
G _{ n \kappa} (r) \right| ^2 + \left|
F _{ n \kappa} (r) \right| ^2  \biggr] \; ,
\end{eqnarray}
and the sum over $n \kappa$ extends over all occupied states.

\begin{figure}
\begin{center}
\resizebox{0.45\textwidth}{!}{\includegraphics{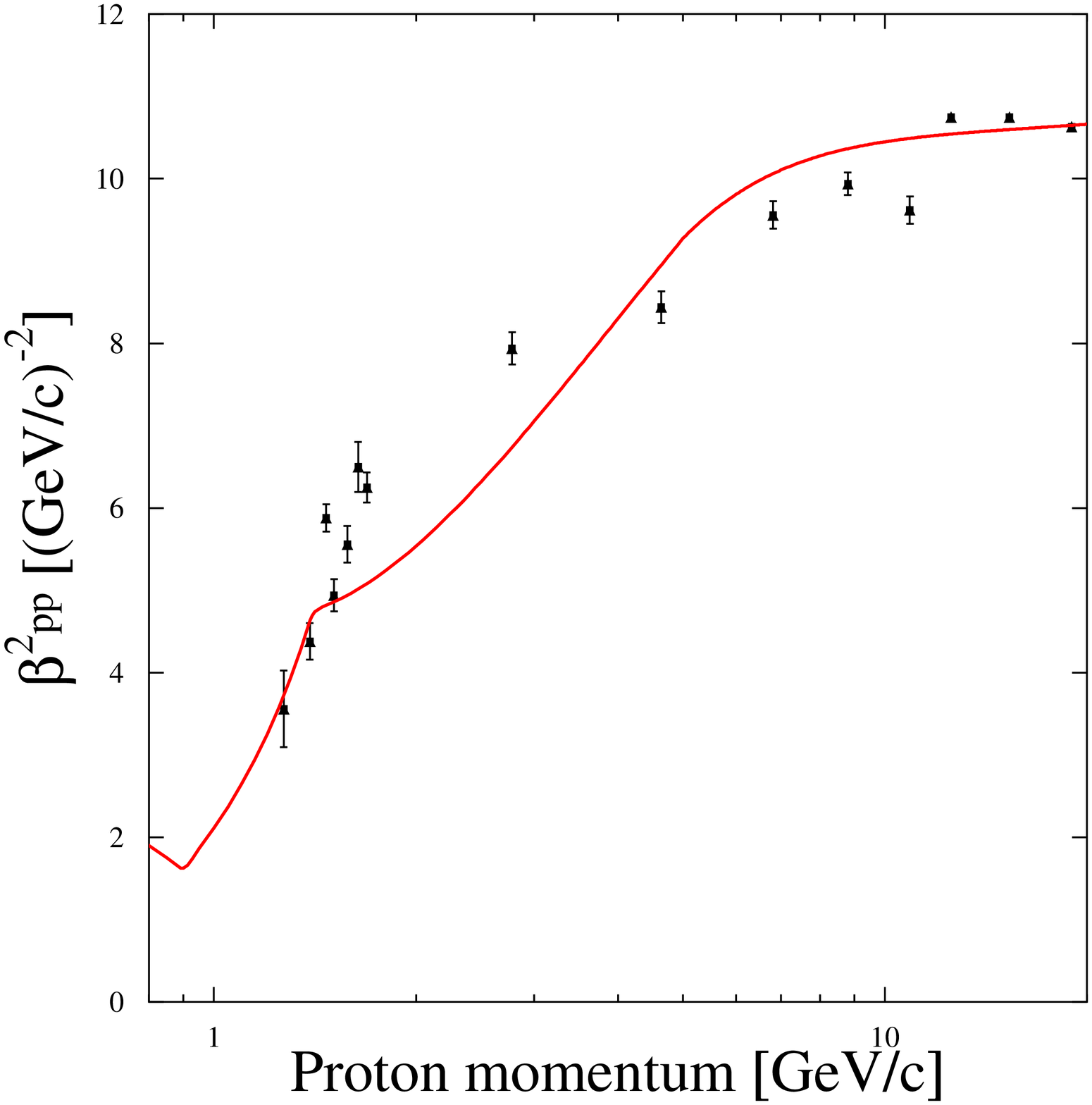}}
\resizebox{0.45\textwidth}{!}{\includegraphics{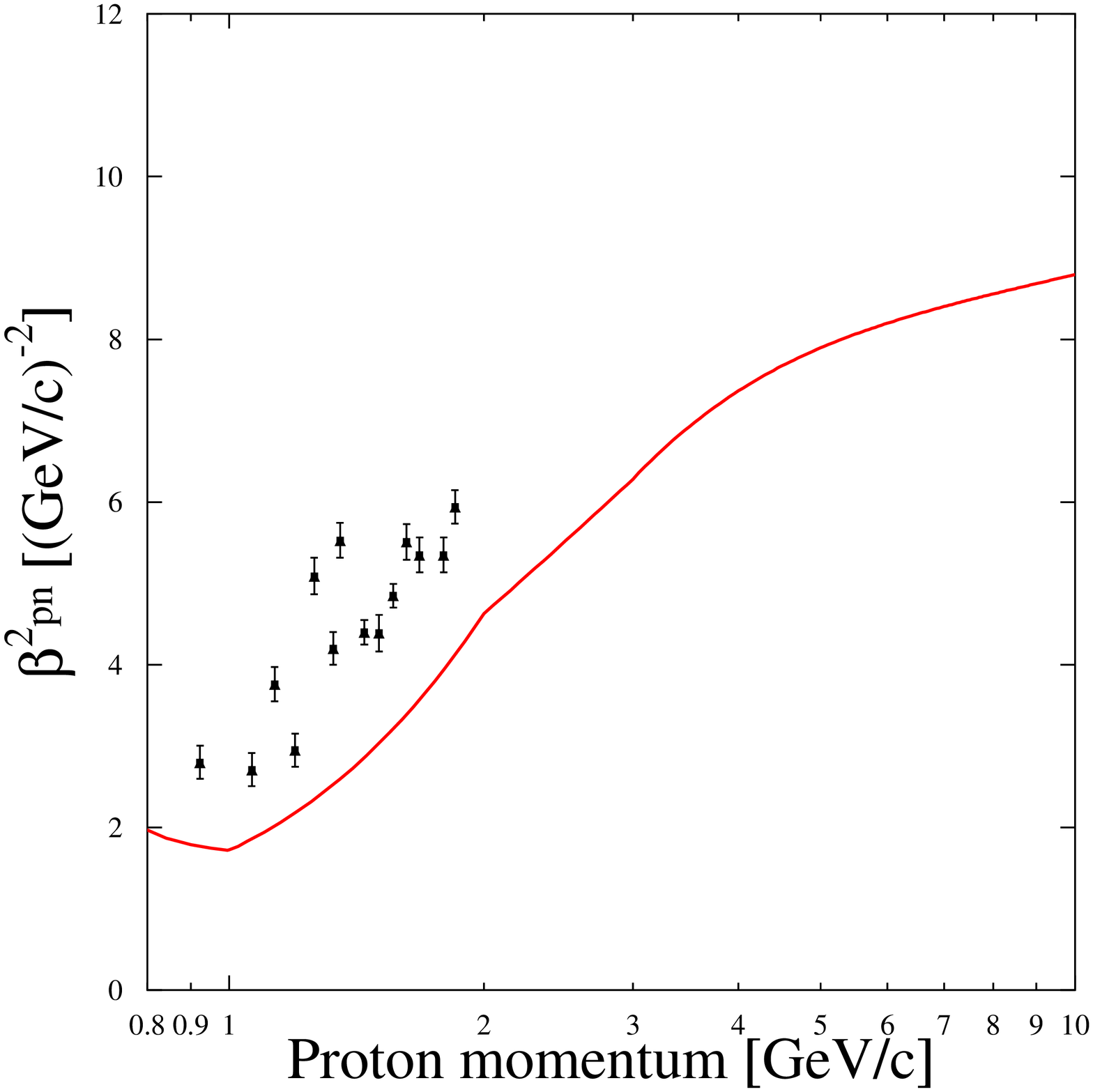}}
\end{center}
\caption{\label{fig:parbeta} The Glauber slope parameters $\beta _{pp}
^2$
and $ \beta _{pn} ^2$ as obtained from Eq.~(\ref{eq:slopeparam}) with the
global fits contained in Fig.~\ref{fig:partotcross}. The data
are from Refs.~\cite{Silverman89} (proton-neutron) and
\cite{Dobrovolsky83} (proton-proton)  and are determined from 
the small-angle $t$ dependence of the measured differential cross sections.}
\end{figure}

\begin{figure}
\begin{center}
\resizebox{0.75\textwidth}{!}{\includegraphics{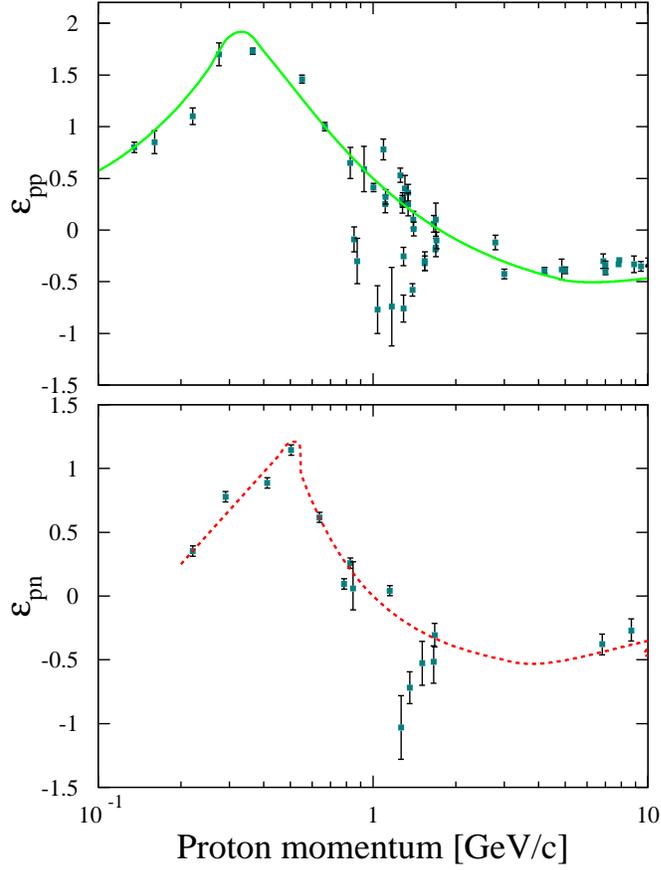}}
\end{center}
\caption{\label{fig:pareps} The ratios of the real to imaginary parts
of the central amplitude $f_{pN}^c$ for proton-proton and proton-neutron
scattering as a function of the proton lab momentum. The curves are
our global fits to the data points. The data are from the review paper
of Ref.~\cite{alkhazov78}.}
\end{figure}

\subsection{\label{sec:born}Glauber parameters}

We refer to $A(e,e'p)$ results obtained with a scattering state of the
form of Eq.~(\ref{eq:glauber}) as a relativistic multiple-scattering
Glauber approximation (RMSGA) calculation.  It is worth stressing that
in contrast to the RDWIA models, all parameters entering the
calculation of the scattering states in the RMSGA $A(e,e'p)$ model can
be directly determined from the elementary proton-proton and
proton-neutron scattering data.  In practice, for a given ejectile's
lab momentum $\mid \vec{k}_p \mid$ the following input is required :
the total proton-proton ${\sigma^{tot}_{pp}}$ and proton-neutron
${\sigma^{tot}_{pn}}$ cross sections, the corresponding slope
parameters (${\beta_{pp}^2}$ and ${\beta_{pn}^2}$) and the ratios of the
real to imaginary part of the scattering amplitude (${\epsilon_{pp}}$
and ${\epsilon_{pn}}$).  We obtain the numbers ${\sigma^{tot}_{pN}}$,
${\beta_{pN} ^2}$ and ${\epsilon_{pN}}$ through interpolation of the data
base available from the Particle Data Group \cite{pdg}.  The slope
parameters $\beta_{pp}^{2}$ and $\beta_{pn}^{2}$ may be found by
analyzing the shape of the differential cross sections assuming that
the contribution from the spin-dependent terms is negligible. At
proton momenta $p_{p} \leq$ 1 GeV the slope parameters found directly
from experiment and phase-shift analysis differ significantly due to a
large contribution from the spin-dependent scattering amplitude
\cite{alkhazov78}.
At higher energies this difference drops quickly indicating that spin
effects are small in that region. In our calculations, the slope
parameters are obtained from the ratio of the elastic $\sigma _{pN}
^{el}$ to the total $\sigma _{pN} ^{tot}$ cross section through the
following relation
\begin{equation}
\beta _{pN} ^{2} \approx  \frac { \left( \sigma ^{tot} _{pN} \right)^2 \left(
\epsilon _{pN} ^2 +1 \right) }
{16 \pi \sigma ^{el} _{pN} } \; . 
\label{eq:slopeparam}  
\end{equation}
In Fig.~\ref{fig:parbeta} we compare the slope parameters obtained
through this formulae with those extracted directly from the $t$
dependence of the differential $pN$ cross sections.  The curves in
Fig.~\ref{fig:parbeta} use the above formulae (\ref{eq:slopeparam})
and our global fits to $\sigma ^{tot} _{pN}$, $\sigma ^{el} _{pN}$ and
$\epsilon _{pN}$ shown in Figs.~\ref{fig:partotcross} and
\ref{fig:pareps}.  For proton-proton scattering the situation emerges
to be very satisfactory.  

\begin{figure}
\resizebox{0.85\textwidth}{!}{\includegraphics{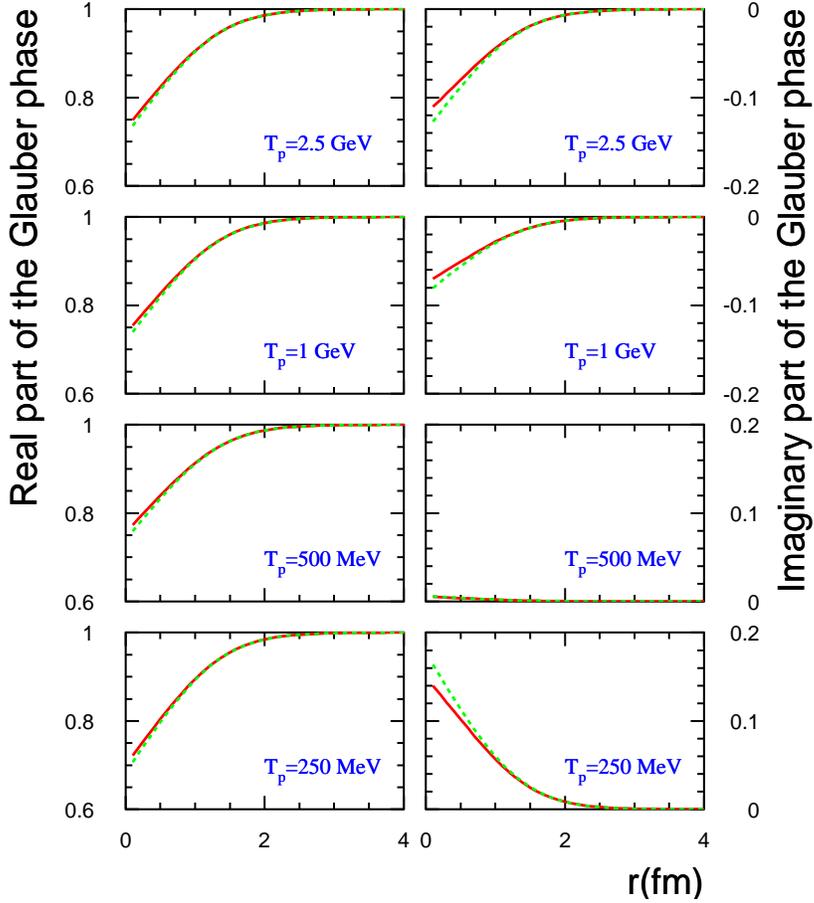}}
\caption{\label{fig:glauphashe} The radial dependence of the real and
imaginary part of the computed Glauber phase $\mathcal{G}$ along the
direction of the ejected particle ( $\theta = 0^ {\circ}$) for proton
emission from $^{4}$He at various proton kinetic energies $T_p$.
Results with the expression of Eq.~(\ref{eq:glaubtrunc}) truncated to
the first order in $\Gamma$ (dashed line) are compared to the full
result (solid line).}
\end{figure}

\begin{figure}
\resizebox{0.85\textwidth}{!}{\includegraphics{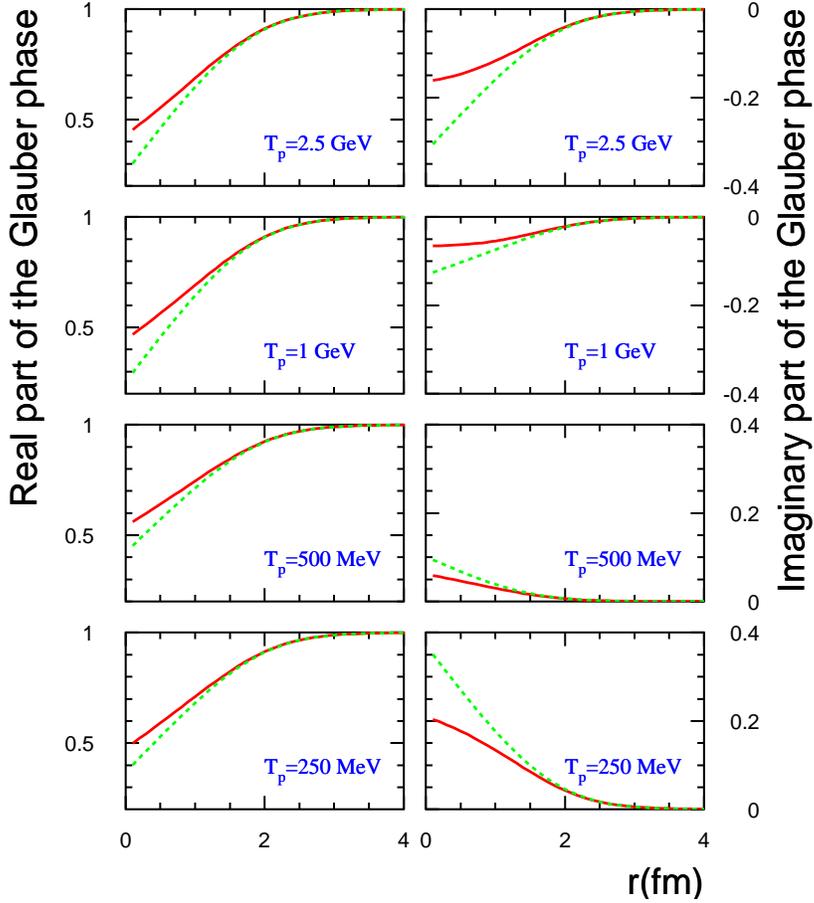}}
\caption{\label{fig:glauphasc} As in Figure~\ref{fig:glauphashe} but
now for $^{12}$C.}
\end{figure}

\begin{figure}
\resizebox{0.85\textwidth}{!}{\includegraphics{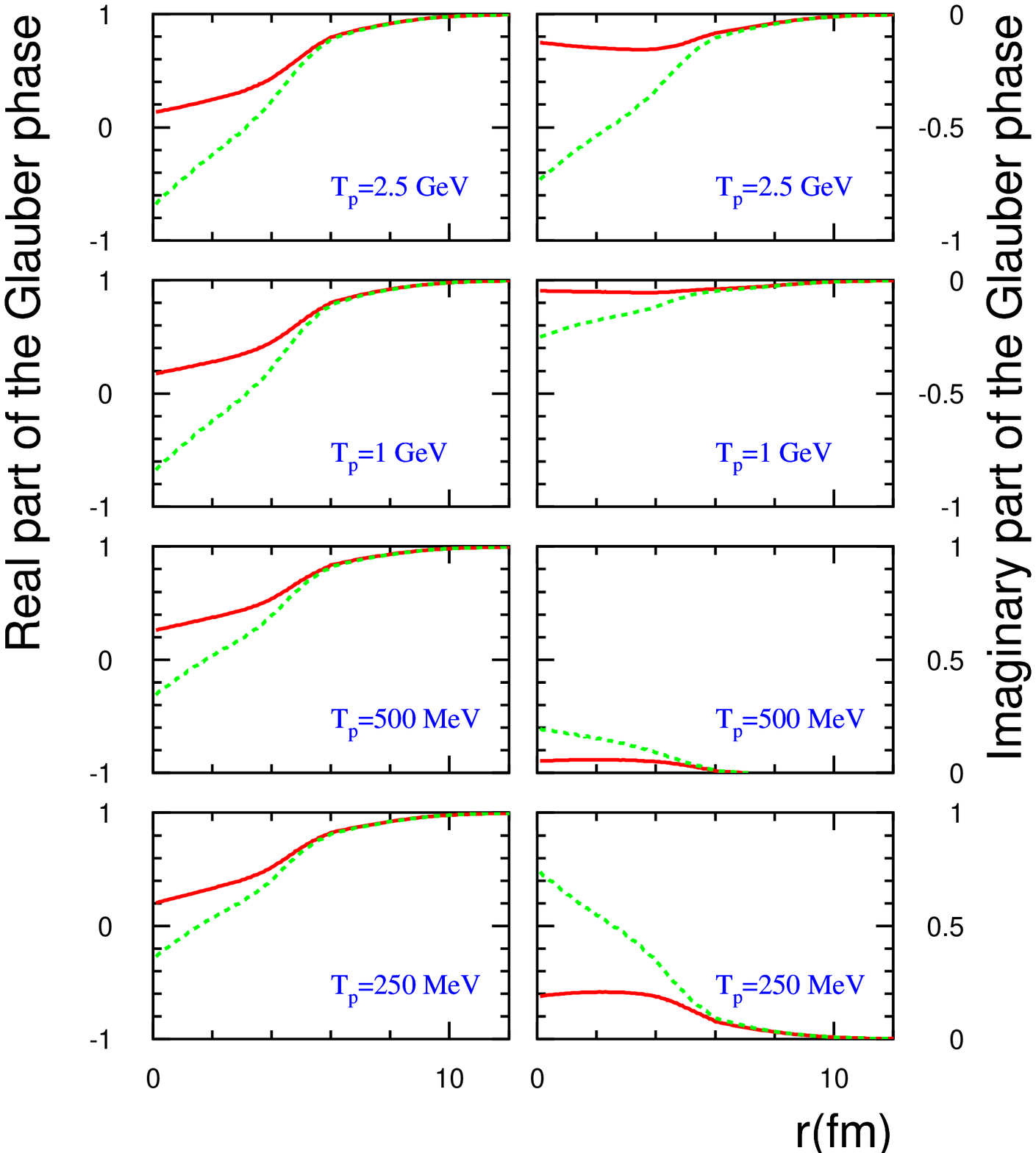}}
\caption{\label{fig:glauphaspb} As in Figure~\ref{fig:glauphashe} but
now for $^{208}$Pb.}
\end{figure}

\section{\label{sec:results} Numerical results for relativistic
Glauber phases}

\subsection{Single- and multiple-scattering effects}
In many works, in the calculation of the Glauber phases only a limited
amount of terms in the expansion of Eq.~(\ref{eq:glaubtrunc}) is
retained.  In some cases, the operator $\mathcal{S}$ is replaced by
the term which is first order in $\Gamma$.  This reduces the treatment
of the FSI effects to one-body operators and corresponds with
retaining the cumulative effect of free passage through the target in
addition to single scatterings. We wish to compare Glauber phases
obtained with the exact operator with those that are produced when
allowing only single scatterings.  The Glauber phase, as it was
defined in Eq.~(\ref{eq:glauberphase}) depends on the two variables
$(b, z) $ and is independent of $\phi _b$.  We wish to remind the
reader that for convenience of the numerical integrations the
$z'$-axis is chosen to point along the asymptotic direction
$\vec{k}_p$ of the ejectile.

In Figs.~\ref{fig:glauphashe}, ~\ref{fig:glauphasc},
\ref{fig:glauphaspb} results are displayed for the computed real and
imaginary part of 
\begin{equation}
\mathcal{G}(r,\theta) = \mathcal{G} \left( b= \sqrt{r^2 - r^2 \cos ^2 \theta} \; ,
z = r \cos \theta) \right) \; ,
\end{equation}
for the target nuclei $^{4}$He, $^{12}$C and $^{208}$Pb and $\theta =
0^o$.  Here, $\theta$ denotes the polar angle with respect to the axis
defined by the asymptotic momentum of the ejected particle.  We remind
that in the absence of final-state interactions the real part of the
Glauber phase equals one, whereas the imaginary part vanishes
identically. As becomes clear from Fig.~\ref{fig:glauphashe}, the
approximation of retaining only single-rescattering terms, emerges to
be a reasonable approximation for the $^4$He target nucleus.  As a
consequence, for $^4$He the average number of rescatterings can be
inferred not to be larger than one. For a given ejectile's momentum,
the average number of scatterers which it encounters in its way out of
the nucleus is expected to grow like $A^{1/3}$.  Given that for $A=4$
the average number of rescatterings is not larger than 1, one can
infer that for a heavy nucleus like $^{208}$Pb the scattering series
of Eq.~(\ref{eq:glaubtrunc}) will receive sizeable contributions up to
the fourth order in $\Gamma$.  This complies with the numbers quoted
in Table~1 of Ref.~\cite{nikolaev95}.  As can be inferred from
Figs.~\ref{fig:glauphashe}, \ref{fig:glauphasc}, \ref{fig:glauphaspb},
single rescatterings dominate the real and imaginary part of the
Glauber phase at the nuclear surface.  In the interior of the nucleus,
single- and the summed higher-order scattering terms come with an
opposite sign for all target nuclei studied.  As a matter of fact,
even for a nucleus like $^{12}$C, the truncation to single scatterings
results in a sizeable overestimation of the FSI effects.  The real
part of the Glauber phase exhibits relatively little $T_p$ dependence
over the range 0.25 - 2.00~GeV covered in the figures.  The imaginary
part, on the other hand, changes sign as one exceeds $T_p=0.5$~GeV and
enters a highly inelastic regime.  This observed change in the
relative sign between the real and imaginary parts of the Glauber
phase is governed by the $T_p$ dependence of the parameters $\epsilon
_{pp}$ and $\epsilon _{pn}$ as is shown in Fig.~\ref{fig:pareps}.

\begin{figure}
\begin{center}
\resizebox{0.85\textwidth}{!}{\includegraphics{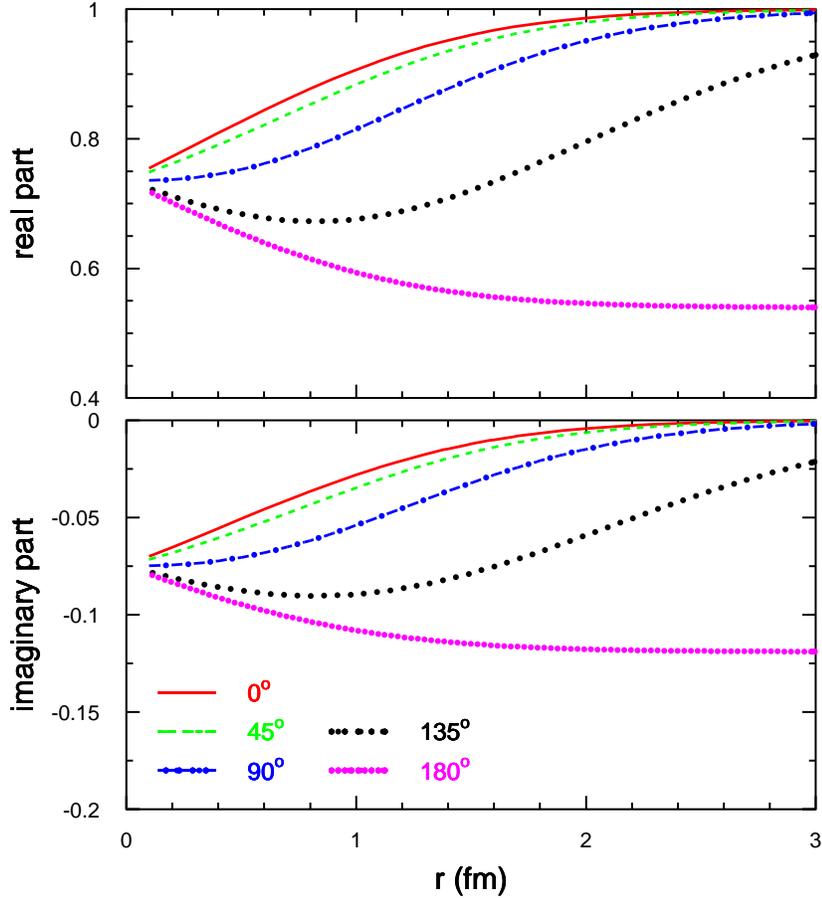}}
\end{center}
\caption{\label{fig:glaudiffangles} The radial dependence of the real and
imaginary part of the computed Glauber phase $\mathcal{G}$ at various
values of $\theta $ and $T_p$=1~GeV for proton emission out of
$^{4}$He.}
\end{figure}

When evaluating the $(e,e'p)$ matrix elements and performing the
integrations over $r$ and $\theta$, the functions displayed in
Figs.~\ref{fig:glauphashe} - \ref{fig:glauphaspb} quantify the effects
stemming from the FSI along the direction defined by the asymptotic
momentum of the ejectile.  For the Glauber phases, the radial
coordinate ``$r$'' can be interpreted as the distance relative to the
center of the target nucleus where the photon hits the target nucleus.
For a given $r$, an additional non-trivial integration over the polar
angles $\theta$ has to be performed.  Here, $0 ^o \le \theta \le 90
^o$ ($90 ^o \le \theta \le 180 ^o$) refers to a situation where the
photon hits the nucleon in the forward (backward) hemisphere with
respect to the direction defined by $\vec{k}_p$.  The dependence of
the real and imaginary part of the Dirac Glauber phase on the polar
angle $\theta$ is illustrated in Fig.~\ref{fig:glaudiffangles} for
emission of 1~GeV protons from $^4$He.  The $\theta = 180^o$ case
corresponds with a peculiar event whereby the photon couples to the
proton along the direction defined by $ - \vec{k} _p$.  For $\theta =
180 ^o$ and increasing $r$, the photon initially hits the proton at
the outskirts of the nucleus and the proton has to travel through the
whole nucleus before it becomes asymptotically free at the opposite
side.  It speaks for itself that these kinematical situations induce
the largest FSI effects but cannot be expected to provide large
contributions to the integrated matrix elements.

\begin{figure}
\begin{center}
\resizebox{0.85\textwidth}{!}{\includegraphics{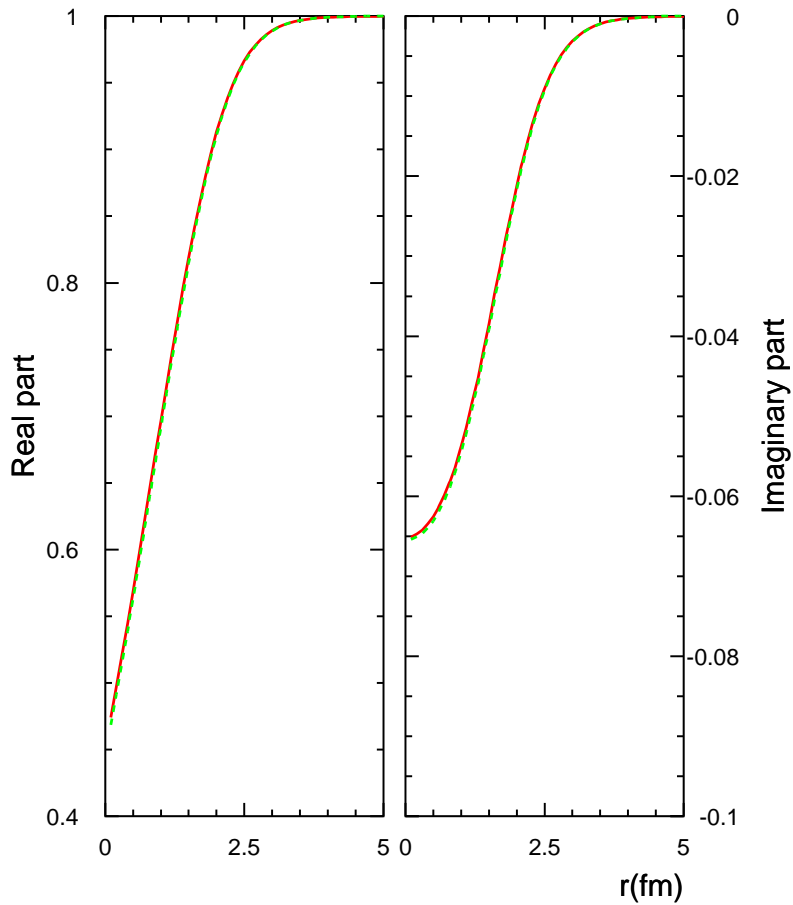}}
\end{center}
\caption{\label{fig:glauphasrel} The effect of the lower components in the
wave functions for the scattering centers on the computed Glauber
phase $\mathcal{G}$ at $\theta = 0^o$ and $T_p =1$~GeV for proton
emission out of $^{12}$C.  Results with ignoring the lower components
(dashed line) are compared to the full result (solid line).}
\end{figure}

\begin{figure}
\begin{center}
\resizebox{0.85\textwidth}{!}{\includegraphics{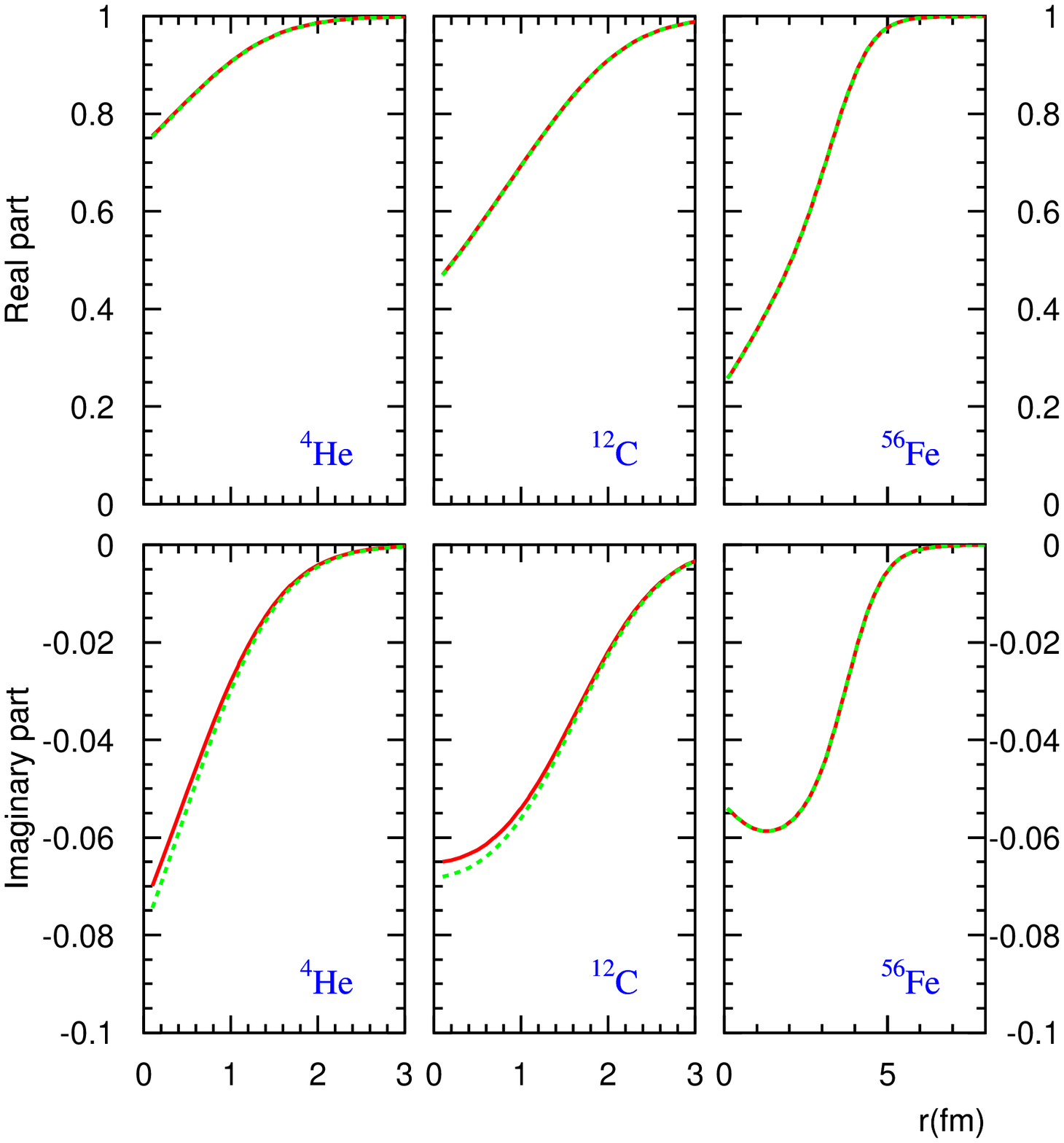}}
\end{center}
\caption{\label{fig:glauphasdens} Effect of replacing the wave
functions for the individual scattering centers by an average density
on the computed Glauber phase $\mathcal{G}$.  The results are obtained
for $\theta = 0^o$ and $T_p =1$~GeV for proton emission out of
$^{4}$He, $^{12}$C and $^{56}$Fe.  Results with the approximated
expression from Eq.~(\ref{eq:glauberphase1}) (dashed line) are
compared to the exact result from Eq. (\ref{eq:glauberphase}) (solid line).}
\end{figure}

\subsection{Relativistic and density effects}

The role of relativity in the description of the FSI can be estimated
by neglecting the small components $F_{n \kappa}(r)$ in the
relativistic wave functions for the individual scattering nucleons in
Eq.~(\ref{eq:glauberphase}) and comparing it to the exact result.  We
have performed several of these calculations for a variety of target
nuclei.  One representative result is displayed in
Fig.~\ref{fig:glauphasrel}.  In general, the relativistic lower wave
function components for the scattering centers (i.e. the nucleons
residing in the daughter nucleus) are observed to have a minor impact
on the predictions for both the real and imaginary part of the Glauber
phase.  In the next section, however, it will be shown that inclusion
of the lower relativistic components is essential for some $A(e,e'p)$
observables.  From the results presented here, it can be excluded that
this could be attributed to a relativistic effect in the description
of the final-state interactions.

Besides the neglect of relativistic effects, most Glauber calculations
use an average nuclear density to describe the spatial distribution
for each of the frozen nucleons from which the ejectile can scatter.
This approximation was outlined at the end of Sect.~
\ref{sec:glaurelA}.  One may naively expect that this ``averaging'' at
the wave-function level becomes increasingly accurate as the target
mass number $A$ increases.  As a matter of fact, the results displayed
in Fig.~\ref{fig:glauphasdens} illustrate that the quoted
approximation is a valid one even for a light nucleus like $^4$He.
Inspecting Fig.~\ref{fig:glauphasdens}, only in the absorptive part for
$^4$He and $^{12}$C a minor overestimation gets introduced through the
averaging procedure.

\begin{figure}
\begin{center}
\resizebox{0.85\textwidth}{!}{\includegraphics{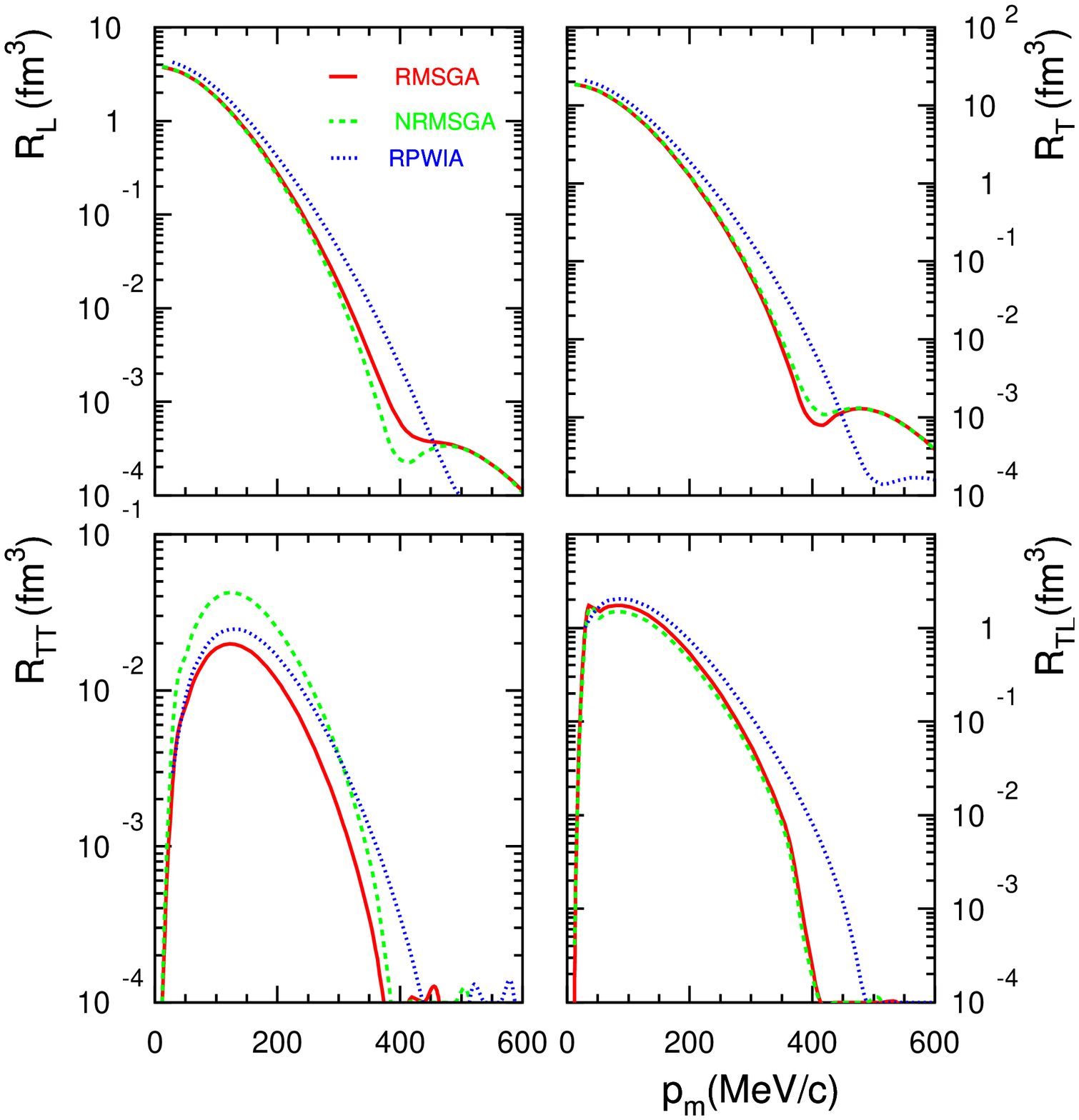}}
\end{center}
\caption{\label{fig:he4qw0.8gev} Separated $^{4}$He$(e,e'p)^{3}$H
response functions versus missing momentum in constant $(q,
\omega)$-kinematics under quasi-elastic conditions. The electron
kinematics was $\epsilon$=4.80~GeV, $q$=1.50~GeV and $\omega$=0.84~GeV
(i.e., $Q^2$=1.54~(GeV)$^2$).}
\end{figure}

\begin{figure}
\begin{center}
\resizebox{0.85\textwidth}{!}{\includegraphics{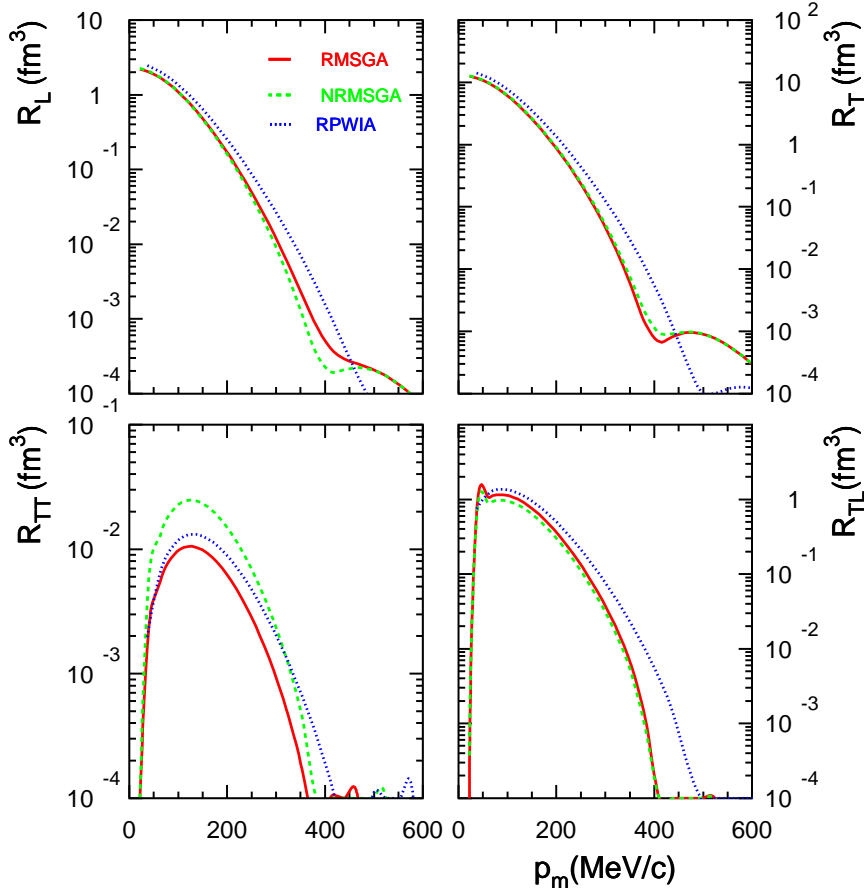}}
\end{center}
\caption{\label{fig:he4qw1gev} Separated $^{4}$He$(e,e'p)^{3}$H
structure functions versus missing momentum in constant $(q,
\omega)$-kinematics under quasi-elastic conditions. The electron
kinematics was $\epsilon$=2.442~GeV, $q$=1.696~GeV and $\omega$=1~GeV
(i.e., $Q^2$=1.88~(GeV)$^2$).}
\end{figure}

\begin{figure}
\begin{center}
\resizebox{0.45\textwidth}{!}{\includegraphics{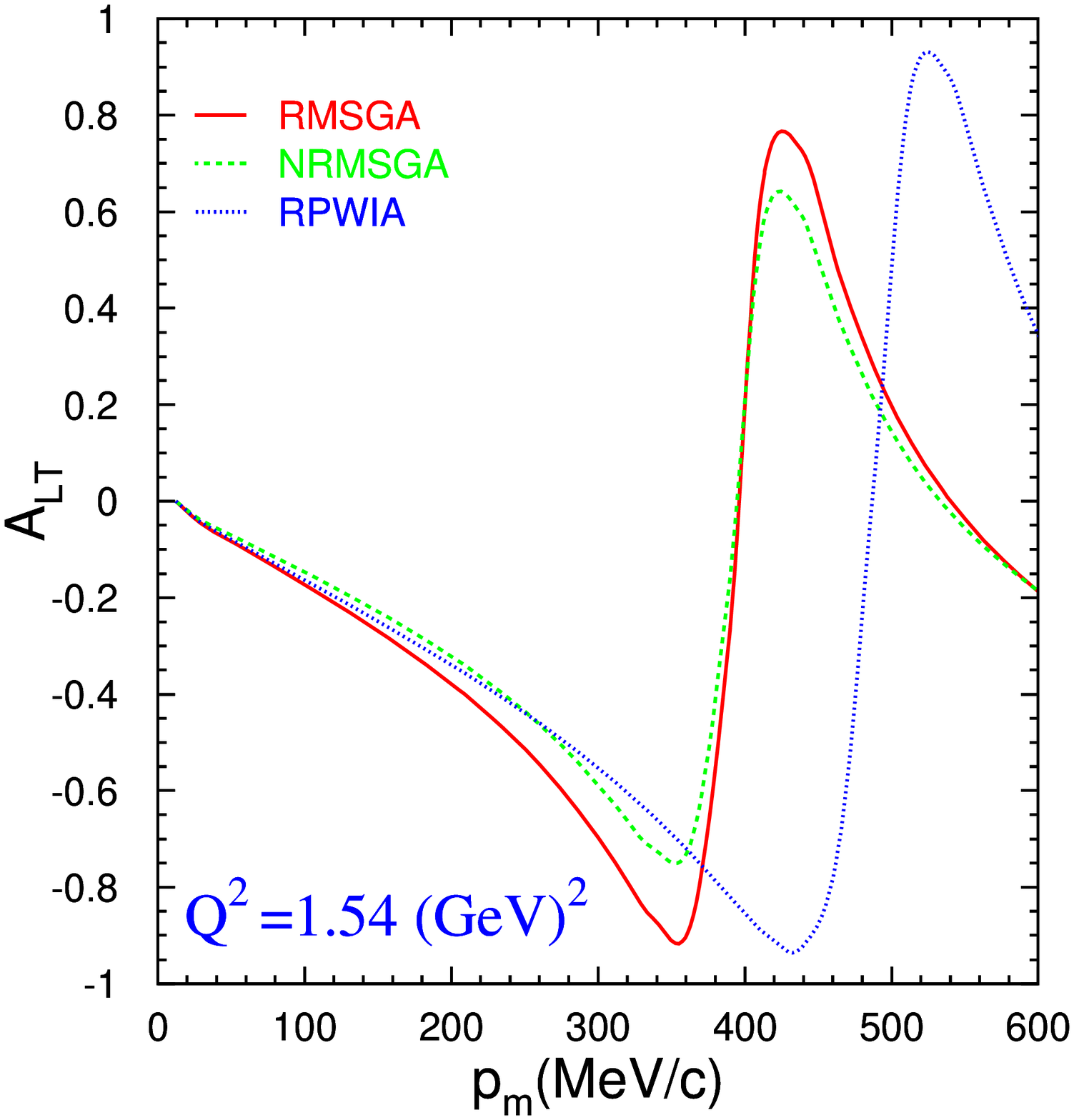}}
\resizebox{0.45\textwidth}{!}{\includegraphics{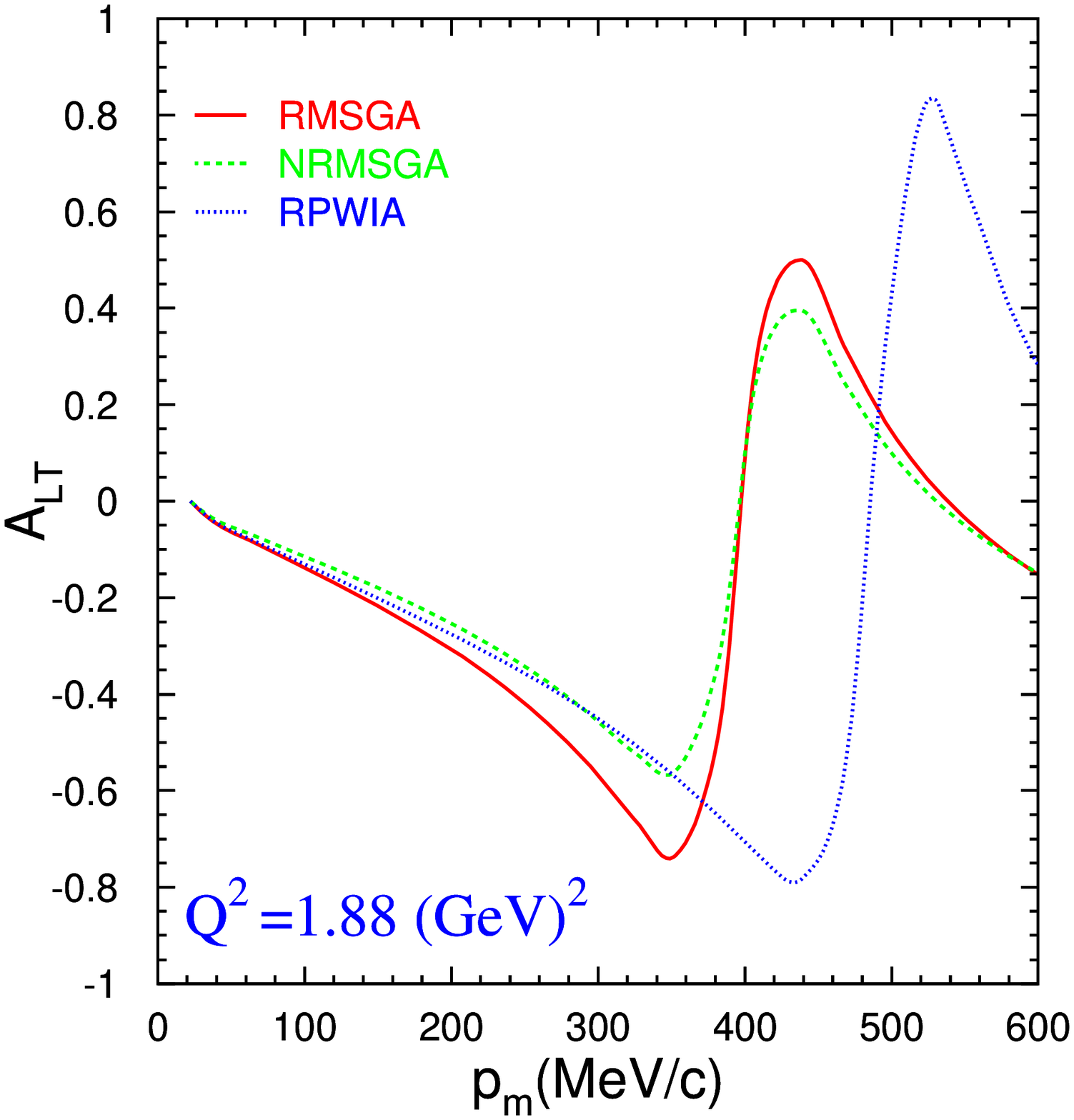}}
\end{center}
\caption{\label{fig:he4alts} The $A_{TL}$ left-right asymmetries as a
function of the missing momentum 
corresponding with the kinematics of Figs.~\ref{fig:he4qw0.8gev} and
\ref{fig:he4qw1gev}. }
\end{figure}

\section{\label{sec:he4resp} Numerical results for $^4$He$(e,e'p)$ structure functions}

In conventional nuclear-physics models, by which we understand models
which are based on hadronic degrees-of-freedom, the $^4$He nucleus
plays a key role.  Indeed, it is the simplest nuclear system in which
all basic characteristics of ``complex'' nuclei exist.  Accordingly,
it comes at no surprise that the $^4$He$(e,e'p)^{3}$H reaction plays a
pivotal role in investigations into the short-range structure of
nuclei \cite{he4expjlab} and medium-modification effects
\cite{dieterich}.  In Figs.~\ref{fig:he4qw0.8gev} and
\ref{fig:he4qw1gev} we display some predictions for the separated
$^4$He$(e,e'p)^{3}$H response functions in kinematics presently
accessible at the Thomas Jefferson National Accelerator Facility. As a
matter of fact, the kinematics corresponding with
Fig.~\ref{fig:he4qw0.8gev} coincides with a scheduled experiment at
this facility \cite{he4expjlab}.

Non-relativistic $A(e,e'p)$ calculations typically miss, amongst other
things, the effect from the coupling between the lower components in
the bound and scattering states.  As a matter of fact, we interpret
the contribution from the coupling between the lower component in the
bound ($\phi _{\alpha _{1}}$) and scattering state ($\phi _{k_{p}
m_s}$) to the matrix element of Eq.~(\ref{eq:glauampfinal}) as a
measure for the impact of relativistic effects.  We denote $A(e,e'p)$
results that are obtained after omitting this specific part as
``NRMSGA''.  We wish to stress that these ``NRMSGA'' calculations
still use relativistic kinematics and bound and scattering states
obtained by solving a Dirac equation.  As can be appreciated from
Figs.~\ref{fig:he4qw0.8gev} and \ref{fig:he4qw1gev}, the effect of the
coupling between the lower components is marginal for the structure
functions $\mathcal{R}_L$ and $\mathcal{R}_T$, but substantial for the
two interference functions $\mathcal{R}_{TT}$ and $\mathcal{R}_{TL}$.
These conclusions regarding the role of relativistic effects at low
missing momenta confirms the major findings of numerous other
investigations \cite{udias00,martinez02}.  Note that the genuine
relativistic effect stemming from the coupling between the lower
components is very prominent in the (non)-filling of the predicted dip
in the cross section for $p_m \approx 410$~ MeV. In order to
illustrate the substantial impact of relativistic effects on some
observables, Fig.~\ref{fig:he4alts} presents the so-called left-right
asymmetry
\begin{equation}
A_{LT} = \frac { \sigma (\phi  = 0^o) - \sigma ( \phi  =
180^o ) }
{ \sigma (\phi  = 0^o) + \sigma ( \phi  =
180^o ) } = \frac { v_{TL} {\mathcal R}_{TL} }
{ \biggl[ v_{L}{\mathcal R}_{L}+v_{T}{\mathcal R}_{T}+v_{TT}{\mathcal
R}_{TT} \biggr] } \; ,
\end{equation}  
corresponding with the kinematics of Figs.~\ref{fig:he4qw0.8gev} and
\ref{fig:he4qw1gev}.  As can be seen the genuine relativistic effect
due to contribution from the the lower components in the bound and
scattering state bring about a substantial increase in $A_{LT}$.  As a
matter of fact, the RMSGA $A_{LT}$ predictions of
Fig.~\ref{fig:he4alts} are completely in line with the
optical-potential RDWIA predictions by J. Udias contained in
Ref.~\cite{he4expjlab}.   

The relativistic plane wave impulse approximation (RPWIA) results of
Figs.~\ref{fig:he4qw0.8gev} - \ref{fig:he4alts} are obtained after
setting the Glauber phase $\mathcal{G} (b,z)$ equal to one in the
matrix elements of Eq.~(\ref{eq:glauampfinal}).  This corresponds with
a calculation which ignores all FSI mechanisms but adopts relativistic
kinematics, Dirac bound states, a relativistic plane wave for the
ejectile and a relativistic current operator.  At low missing momentum
the FSI quenches the cross sections by 20 \%, confirming the result of
the non-relativistic Glauber calculations of Refs.~\cite{benhar00},
\cite{Morita2002} and the calculations of Laget reported in
Ref.~\cite{Laget94}.  Not surprisingly, with growing missing momentum
the FSI effects become increasingly important.  For example the dip in
the RPWIA response functions at $p_m \approx 0.5$~GeV is shifted
downwards by about 100~MeV and partially washed out.

\section{\label{sec:concl} Summary and conclusions}

A fully unfactorized relativistic formulation of Glauber
multiple-scattering theory for modeling exclusive $A(e,e'p)$ processes
has been presented.  Formally, the model bears a strong resemblance
with the RDWIA approaches which have been developed over the last
number of decades.  In contrast to the RDWIA models, the relativistic
Glauber approximation for dealing with FSI mechanisms holds stronger
links with the elementary proton-proton and proton-neutron processes
and does not require the (phenomenological) input of optical
potentials.  Our fully unfactorized framework can accommodate all
relativistic effects which are usually implemented in the RDWIA
approaches.  Like in the RDWIA models, the bound-state wave functions
are solutions to a Dirac equation with scalar and vector potentials
fitted to the ground-state nuclear properties, i.e. an approach
commonly known as the $\sigma - \omega$ model.

The relative contribution from single- and multiple-scatterings to the
integrated FSI mechanisms, has been investigated.  For the target
nucleus $^4$He, it turned out that single-scatterings give already a
fair account of the complete scattering processes, and that double-
and triple-scatterings are rare.  For target nuclei heavier than
$^4$He, the net effect of double-, triple- and quadruple-scatterings
is found to affect the real and imaginary parts of the FSI induced
phase with an opposite sign as compared to the single-scattering term.
For $^{12}$C and heavier target nuclei, Glauber calculations 
restricted to single-scatterings substantially overestimate the FSI
mechanisms.  The lower components in the relativistic single-particle
wave functions, on the other hand, are observed to have a negligible
impact on the predictions for the distorting and absorptive effect of
single- and multiple-scattering events which the ejected proton
undergoes. This observation provides support for the non-relativistic
Glauber approaches which have been widely adopted in nuclear
transparency calculations for example. Also the frequently adopted
simplification of replacing the squared nucleon wave functions for the
individual scatterering centers in the target nucleus by some averaged
nuclear density, was found to lead to results which approximate nicely
the exact ones.

The developed RMSGA model has been applied to the exclusive $^4$He $+e
\longrightarrow e' + p + t $ process.  The predicted RMSGA structure
functions are found to follow similar trends in comparison to
predictions from non-relativistic Glauber and relativistic 
optical-potential calculations.  Extensions of the presented model
include the implementation of short-range nucleon-nucleon correlations
and color transparency effects.  Work in this direction is in progress.

%\begin{acknowledgments}
%\end{acknowledgments} 

%\bibliographystyle{elsart-num}
%\bibliography{/home/jan/papers/myreferences}

\begin{thebibliography}{10}
\expandafter\ifx\csname url\endcsname\relax
  \def\url#1{\texttt{#1}}\fi
\expandafter\ifx\csname urlprefix\endcsname\relax\def\urlprefix{URL }\fi

\bibitem{pandharipande97}
V.~Pandharipande, I.~Sick, P.~deWitt Huberts, Rev. Mod. Phys. 69 (1997) 981.

\bibitem{muether02}
H.~M\"{u}ther, A.~Polls, Prog. Part. Nucl. Phys. 45 (2000) 243.

\bibitem{gao00}
J.~{Gao et al.}, Phys. Rev. Lett. 84 (2000) 3265.

\bibitem{malov00}
S.~{Malov et al.}, Phys. Rev. C 62 (2000) 057302.

\bibitem{dieterich}
S.~{Dieterich et al.}, Phys. Lett. B 500 (2001) 47.

\bibitem{Garrow02}
K.~{Garrow et al.}, Phys. Rev. C 66 (2002) 044613.

\bibitem{boffi93}
S.~Boffi, C.~Giusti, F.~Pacati, Phys. Rep. 226 (1993) 1.

\bibitem{Kelly96}
J.~Kelly, Adv. Nucl. Phys 23 (1996) 75.

\bibitem{picklesimer85}
A.~Picklesimer, J.~{Van Orden}, S.~Wallace, Phys. Rev. C 32 (1985) 1312.

\bibitem{johansson96}
J.~Johansson, H.~Sherif, G.~Lotz, Nucl. Phys. A 605 (1996) 517.

\bibitem{yin92}
Y.~Yin, D.~Onley, L.~Wright, Phys. Rev. C 45 (1992) 1311.

\bibitem{udias93}
J.~Udias, P.~Sarriguren, E.~{Moya de Guerra}, E.~Garrido, J.~Caballero, Phys.
  Rev. C 48 (1993) 2731.

\bibitem{Meucci2000}
A.~Meucci, C.~Giusti, F.~Pacati, Phys. Rev. C 64 (2000) 64014605.

\bibitem{serot86}
B.~Serot, J.~Walecka, Adv. Nucl. Phys. 16 (1986) 1.

\bibitem{udias00}
J.~Udias, J.~Vignote, Phys. Rev. C 62 (2000) 034302.

\bibitem{martinez02}
M.~Mart\'{i}nez, J.~Caballero, T.~Donnelly, Nucl. Phys. A 707 (2002) 83.

\bibitem{glauber70}
R.~Glauber, G.~Matthiae, Nucl. Phys. B 21 (1970) 135.

\bibitem{Wallace75} S.~J. Wallace, Phys.  Rev. C 12 (1975) 179.

\bibitem{Yennie71}
D.~Yennie, Interaction of high-energy photons with nuclei as a test of
  vector-meson-dominance, in: J.~Cummings, D.~Osborn (Eds.), Hadronic
  Interactions of Electrons and Photons, Academic, New York, 1971, p. 321.

\bibitem{benhar00}
O.~Benhar, N.~Nikolaev, J.~Speth, A.~Usmani, B.~Zakharov, Nucl. Phys. A 673
  (2000) 241.

\bibitem{Claudio01}
C.~{Ciofi degli Atti}, L.~Kaptari, D.~Treleani, Phys. Rev. C 63 (2001) 044601.

\bibitem{jeschonnek99}
S.~Jeschonnek, T.~Donnelly, Phys. Rev. C 59 (1999) 2676.

\bibitem{kohama00}
A.~Kohama, K.~Yazaki, R.~Seki, Nucl. Phys. A 662 (2000) 175.

\bibitem{frankfurt95}
L.~Frankfurt, E.~Moniz, M.~Sargsyan, M.~Strikman, Phys. Rev. C 51 (1995) 3435.

\bibitem{nikolaev95}
N.~Nikolaev, A.~Szcurek, J.~Speth, J.~Wambach, B.~Zakharov, V.~Zoller, Nucl. Phys. A 582 (1995) 665.

\bibitem{Petraki2003} M.~Petraki, E.~Mavrommatis, O.~Benhar, J.~Clark,
A.~Fabrocini, S.~Fantoni, Phys. Rev. C 67 (2003) 014605.

\bibitem{amado83}
R.~Amado, J.~Piekarewicz, D.~Sparrow, J.~McNeil, Phys. Rev. C 28 (1983) 1663.

\bibitem{greenberg94}
W.~Greenberg, G.~Miller, Phys. Rev. C 49 (1994) 2747.

\bibitem{debruyne00}
D.~Debruyne, J.~Ryckebusch, W.~{Van Nespen}, S.~Janssen, Phys. Rev. C 62 (2000)
  024611.

\bibitem{donnelly86}
T.~Donnelly, A.~Raskin, Ann. Phys. 169 (1986) 247.

\bibitem{raskin89}
A.~Raskin, T.~Donnelly, Ann. Phys. 191 (1989) 78.

\bibitem{bjorken64}
J.~Bjorken, S.~Drell, Relativistic Quantum Mechanics, Mc-Graw-Hill, New York,
  1964.

\bibitem{cooper93}
E.~Cooper, S.~Hama, B.~Clarck, R.~Mercer, Phys. Rev. C 47 (1993) 297.

\bibitem{ito97}
H.~Ito, S.~Koonin, R.~Seki, Phys. Rev. C 56 (1997) 3231.

\bibitem{bianconi95}
A.~Bianconi, M.~Radici, Phys. Lett. B 363 (1995) 24.

\bibitem{bianconi96}
A.~Bianconi, M.~Radici, Phys. Rev. C 54 (1996) 3117.

\bibitem{alkhazov78}
G.~Alkhazov, S.~Belostotsky, A.~Voroboyov, Phys. Rep. 42 (1978) 89.

\bibitem{pdg}
K.~{Hagiwara}, K.~{Hikasa}, K.~{Nakamura}, M.~{Tanabashi},
  M.~{Aguilar-Benitez}, C.~{Amsler}, R.~{Barnett}, P.~{Burchat}, C.~{Carone},
  C.~{Caso}, G.~{Conforto}, O.~{Dahl}, M.~{Doser}, S.~{Eidelman}, J.~{Feng},
  L.~{Gibbons}, M.~{Goodman}, C.~{Grab}, D.~{Groom}, A.~{Gurtu}, K.~{Hayes},
  J.~{Hern\'andez-Rey}, K.~{Honscheid}, C.~{Kolda}, M.~{Mangano}, D.~{Manley},
  A.~{Manohar}, J.~{March-Russell}, A.~{Masoni}, R.~{Miquel}, K.~{M\"onig},
  H.~{Murayama}, S.~{Navas}, K.~{Olive}, L.~{Pape}, C.~{Patrignani},
  A.~{Piepke}, M.~{Roos}, J.~{Terning}, N.~{T\"ornqvist}, T.~{Trippe},
  P.~{Vogel}, C.~{Wohl}, R.~{Workman}, W.-M. {Yao}, B.~{Armstrong}, P.~{Gee},
  K.~{Lugovsky}, S.~{Lugovsky}, V.~{Lugovsky}, M.~{Artuso}, D.~{Asner},
  K.~{Babu}, E.~{Barberio}, M.~{Battaglia}, H.~{Bichsel}, O.~{Biebel},
  P.~{Bloch}, R.~{Cahn}, A.~{Cattai}, R.~{Chivukula}, R.~{Cousins}, G.~{Cowan},
  T.~{Damour}, K.~{Desler}, R.~{Donahue}, D.~{Edwards}, V.~{Elvira},
  J.~{Erler}, V.~{Ezhela}, A.~{Fass\`o}, W.~{Fetscher}, B.~{Fields},
  B.~{Foster}, D.~{Froidevaux}, M.~{Fukugita}, T.~{Gaisser}, L.~{Garren}, H.-J.
  {Gerber}, F.~{Gilman}, H.~{Haber}, C.~{Hagmann}, J.~{Hewett},
  I.~{Hinchliffe}, C.~{Hogan}, G.~{H\"ohler}, P.~{Igo-Kemenes}, J.~{Jackson},
  K.~{Johnson}, D.~{Karlen}, B.~{Kayser}, S.~{Klein}, K.~{Kleinknecht},
  I.~{Knowles}, P.~{Kreitz}, Y.~{Kuyanov}, R.~{Landua}, P.~{Langacker},
  L.~{Littenberg}, A.~{Martin}, T.~{Nakada}, M.~{Narain}, P.~{Nason},
  J.~{Peacock}, H.~{Quinn}, S.~{Raby}, G.~{Raffelt}, E.~{Razuvaev}, B.~{Renk},
  L.~{Rolandi}, M.~{Ronan}, L.~{Rosenberg}, C.~{Sachrajda}, A.~{Sanda},
  S.~{Sarkar}, M.~{Schmitt}, O.~{Schneider}, D.~{Scott}, W.~{Seligman},
  M.~{Shaevitz}, T.~{Sj\"ostrand}, G.~{Smoot}, S.~{Spanier}, H.~{Spieler},
  N.~{Spooner}, M.~{Srednicki}, A.~{Stahl}, T.~{Stanev}, M.~{Suzuki},
  N.~{Tkachenko}, G.~{Valencia}, K.~{van Bibber}, M.~{Vincter}, D.~{Ward},
  B.~{Webber}, M.~{Whalley}, L.~{Wolfenstein}, J.~{Womersley}, C.~{Woody},
  O.~{Zenin}, Review of particle physics, Physical Review D 66 (2002) 010001+.
\newline\urlprefix\url{http://pdg.lbl.gov}

\bibitem{Walecka2001}
J.~D. Walecka, Electron Scattering for Nuclear and Nucleon Structure, Cambridge
  University Press, Cambridge, 2001.

\bibitem{varga2002}
K.~Varga, S.~Pieper, Y.~Suzuki, R.~Wiringa, Phys. Rev. C 66 (2002) 034611.

\bibitem{Silverman89}
B.~{Silverman}, J.~{Lugol}, J.~{Saudinos}, Y.~{Terrien}, F.~{Wellers},
  A.~{Dobrolvolsky}, A.~{Khanzadeev}, G.~{Korolev}, G.~{Petrov},
  E.~{Spiridenkov}, A.~{Vorobyov}, Nucl. Phys. A 499 (1989) 763.

\bibitem{Dobrovolsky83}
A.~{Dobrovolsky}, A.~{Khanzadeev}, G.~{Korolev}, E.~{Maev}, V.~{Medvedev},
  G.~{Sokolov}, N.~{Terentyev}, Y.~{Terrien}, G.~{Velichko}, A.~{Vorobyov},
  Y.~{Zalite}, 
  Nucl. Phys. B 214 (1983) 1--20.

\bibitem{he4expjlab}
K.~Aniol, S.~Gilad, D.~Higinbotham, A.~Saha, Detailed study of $^4$he nuclei
  through response function separations at high momentum transfers, Tech. rep.,
  JLAB experiment E-01-020 (2001).
\newline\urlprefix\url{http://www.jlab.org/exp-prog/proposals/01/PR01-20.pdf}

\bibitem{Morita2002}
H.~Morita, M.~Braun, C.~C. degli Atti, D.~Treleani, Nucl. Phys. A A699 (2002) 328c.

\bibitem{Laget94}
J.-M. Laget, Nucl. Phys. A A579 (1994) 333.

\end{thebibliography}

\end{document}